# A Unified Micro-Model for Loss Reserves, IBNR and Unearned Premium Risk with Dependence, Inflation, and Discounting

Emmanuel Hamel[1], Anas Abdallah[2] and Ghislain Léveillé[3]

**Abstract**

This paper introduces a unified micro-level stochastic framework for the joint modeling of loss reserves (RBNS), incurred but not reported (IBNR) reserves, and unearned premium risk under dependence, inflation, and discounting. The proposed framework accommodates interactions between indemnities, expenses, reporting delays, and settlement delays, while allowing for flexible parametric dependence structures and dynamic financial adjustments. An Aggregate Trend Renewal Process (ATRP) is used as one possible implementation of the joint model for payments, expenses, and delays; however, the methodological contribution of the paper lies in the unified micro-level reserving architecture rather than in the ATRP itself. The framework produces forward-looking reserve and premium risk measures with direct applications to pricing, reserving, and capital management.

We implement the framework using an aggregate trend renewal process at the individual claim level, which can be applied to the usual run-off triangle to obtain predictions for each accident–development year. Closed-form expressions for the first two raw and joint conditional moments of predicted payments are derived, together with approximations of their distribution functions. A detailed case study on medical malpractice insurance illustrates the practical relevance of the approach and its calibration on real-world data. We also investigate data heterogeneity, parameter uncertainty, distributional approximations, premium risk, UPR sensitivity to operational delays and inflation, and risk capital implications under alternative assumptions. The results highlight the advantages of unified micro-level modeling for dynamic liability and premium risk assessment in long-tailed lines of business.

*Keywords*: Unearned premium reserve; IBNR; Inflation and discount factors; Dependence modeling; Indemnities and expenses; Medical malpractice; Florida dataset; Risk capital.

## 1. INTRODUCTION

Provisions for payment obligations from losses that have occurred but have not yet been settled usually comprise most of an insurance company's liabilities (see Grace and Leverty (2010)). Therefore, determining and evaluating loss reserving techniques is vital and one of the most critical problems in the insurance industry.

### 1.1. Loss reserving

---

[1] École d'actuariat, Pavillon Paul Comtois, Université Laval, Québec, Canada, G1V 0A6

[2] Corresponding author: Department of Mathematics and Statistics, McMaster University, Hamilton, Ontario, Canada, L8S 4L8, E-mail : anasabdallah@mcmaster.ca

[3] École d'actuariat, Pavillon Paul Comtois, Université Laval, Québec, Canada, G1V 0A6,

Historically, the claims reserves were always based on aggregate claims data structured in claims development triangles, also called run-off triangles. At first, deterministic methods were suggested, and the most used ones on aggregate claims data are the Chain-Ladder and the Bornhuetter & Ferguson methods (see Bornhuetter & Ferguson (1972)). With the advent of the new regulatory standards (e.g. Solvency II in Europe and the ORSA guidelines in North America), the use of stochastic claims reserving methods became increasingly popular, see Wüthrich and Merz (2008) and England and Verrall (2002) for good reviews of the existing techniques therein. Moreover, these new regulations also require modeling dependence between the different components of the portfolio. Non-parametric approaches (see, for example, Braun (2004), Schmidt (2006)) and parametric approaches (see, for example, Shi and Frees (2011), Wüthrich et al. (2013), and Abdallah (2016)) have extensively studied different types of dependence between run-off triangles in an insurance company.

Some macro-reserving studies, such as de Jong (2012), Shi et al. (2012), and Abdallah et al. (2015), focus on modeling dependencies between loss triangles and accommodating inflation at the aggregate level. They often use copulas or hierarchical models to capture correlations between loss triangles or calendar years. In contrast, our ATRP model operates at the micro level, incorporating dependencies directly between random variables like payments, delays, and expenses. This enables a more granular integration of inflation, discount factors, and trend effects. Our approach complements these macro methods by providing closed-form expressions for key moments and predictive distributions, offering enhanced insights for individual claim reserving and risk capital analysis.

Even now, macro-level reserving is considered the benchmark method and has provided consistent and robust results for stable lines of business and homogeneous sub-groups of risks. However, with new technologies and the possibility of using more precise and granular data, the micro-level reserving method has opened a new dimension to researchers and practitioners. Consequently, recent research on loss reserving advocates for micro-level (granular) reserving methods that evaluate the ultimate cost for each claim, see for example, Lopez et al. (2020), Antonio and Plat (2014), Pigeon et al. (2013), and Taylor et al. (2008).

In fact, unlike micro-level reserving methods, the macro-level reserving models do not allow, for example, the incorporation of detailed information about the individual claims behaviour and cannot separate the estimation of the incurred but not reported (IBNR) claims and the reported but not settled (RBNS) claims since aggregate claims data in each "accident" year are used in the estimation. Furthermore, if we add discount and inflation factors, the macro-level methods are certainly less accurate since we must decide (arbitrarily) when these factors should be applied. Several studies have compared the two loss-reserving approaches; see, for example, Hiabu (2017), Charpentier and Pigeon (2016), Huang et al. (2016) and Jin and Frees (2013). In this work, we focus on micro-level modeling of RBNS liabilities within a unified stochastic reserving framework that simultaneously incorporates dependence, inflation, discounting, and operational delays. The general framework itself is not tied to a specific point process specification. For the purposes of illustration and tractability, we implement the framework using an Aggregate Trend Renewal Process (ATRP), which offers analytical convenience and flexibility, but other arrival and delay processes could be used without altering the structure of the framework or its reserving implications.

More recently, Yang et al. (2024) proposed a copula-based point process framework for individual loss reserving, emphasizing the dynamic prediction of outstanding liabilities by capturing dependencies among payment events, amounts, and settlement times. While their approach focuses on recurrent payment processes and dependencies within claim cash flows, our ATRP model differs by providing closed-form expressions for key moments and incorporating practical considerations such as inflation, discount factors, and sensitivity analysis. Both models address important aspects of individual loss reserving and approach the problem from complementary perspectives.

## 1.2. Aggregate trend renewal model for loss reserving

We emphasize that the Aggregate Trend Renewal Process (ATRP) is not the central methodological contribution of this paper. Rather, it serves as a convenient and flexible stochastic mechanism through which we implement and illustrate the broader unified micro-level reserving framework introduced in this study. The key contribution lies in the integration of RBNS, IBNR, and UPR dynamics with dependence, inflation, and discounting within a single coherent micro-level architecture. The ATRP is employed here because of its analytical tractability and its ability to jointly accommodate trend, delay distributions, and payment dynamics, but the framework itself remains agnostic to the specific renewal-type specification.

In this paper, we rely on the same renewal-based model introduced in Léveillé and Hamel (2017, 2019) but adapt it to a loss-reserving setting where granular RBNS information is used to predict future liabilities. In their earlier work, the authors develop a general counting process—introduced by Lindqvist (2003) and referred to as a trend renewal model—that unifies several well-known processes such as the renewal process (RP) and the non-homogeneous Poisson process (NHPP). Here, we make use of the *same* underlying trend renewal structure for all components of the reserving problem. However, for the RBNS portion, we employ a *conditional* version of the model to incorporate observed reporting and settlement information, whereas for the IBNR and UPR components, the model is used in its non-conditional form. This approach yields a coherent and internally consistent modeling framework across RBNS, IBNR, and UPR, while introducing the conditioning only where it is required by the available information.

After conditioning on the information contained in the RBNS losses, our unified micro-level model is used to predict future granular payments while introducing several new tools to the loss-reserving literature. These developments complement the preceding papers of Léveillé and Hamel (2017, 2019), but extend them in a fundamentally different direction: our framework simultaneously models RBNS, IBNR, and UPR within a single coherent structure. The key additions of this paper include the introduction of trend effects on different operational delays, the explicit separation of indemnity and expense payments, and the incorporation of discount and inflation rates that can differ across payment types. Beyond these practical enhancements, the model provides analytic formulas for moments, approximations for reserve distributions, and an integrated treatment of premium risk and IBNR estimation. Building on these foundations, our work introduces several methodological and practical advancements to the reserving literature.

The main contributions of this paper may be summarized as follows.

(1) We develop a unified micro-level stochastic reserving framework that jointly models RBNS, IBNR, and UPR liabilities within a single coherent architecture. This integration ensures internal consistency across all liability components, including their operational delays and financial adjustments.

(2) We introduce the use of an Aggregate Trend Renewal Process (ATRP) as an implementation device for granular loss reserving. While ATRP has appeared in the broader counting process literature, its application to micro-level reserving has not been previously explored. We show how its trend modulation, renewal structure, and analytical tractability make it particularly well suited for illustrating the proposed framework, while noting that alternative arrival or delay mechanisms could also be accommodated.

(3) We explicitly separate indemnity and expense payments and model their dependence jointly with reporting and settlement delays. This distinction is crucial in long-tailed business lines, where indemnity and expense exhibit different operational and financial behaviours.

(4) We propose a modeling structure that jointly incorporates dependence, inflation, and discounting at the micro level. This enables forward-looking prediction of payment streams and reserve distributions, supports sensitivity analyses to operational and financial assumptions, and produces outputs directly relevant for pricing, reserving, and capital management.

Taken together, these contributions advance the micro-level loss reserving literature by providing a flexible, interpretable, and analytically transparent framework capable of capturing the multifaceted dynamics underlying RBNS, IBNR, and UPR estimation. The remainder of the paper implements these contributions within the proposed micro-level architecture and evaluates their implications through analytical derivations, empirical calibration, and sensitivity analyses.

Thus, to achieve these goals, we present in Section 2 the (conditional) ATRP model that will be used to make our (cell and total) predictions with the help of the first two corresponding moments. In Section 3, we present the dataset on which we apply our micro-model, analyze its heterogeneity, calibrate our model on this dataset, compute the first two moments of our predictions, make an approximation of the reserve distribution and examine the parameters uncertainty. In Section 4, we study the impacts on the reserve if some changes are made in the assumptions of our model, as well as model performance and validation. Section 5 delves into unearned premium risk insights, incurred but not reported (IBNR) reserves, and trend analysis, examining the impact of the trend analysis on actuarial practice. Section 6 focuses on IBNR and UPR and premium risk, providing a novel sensitivity analysis of IBNR and UPR to operational delays, inflation, and trend dynamics, with insights for pricing and reserving. In Section 7, a risk capital analysis is made over different hypotheses. In Section 8, we discuss the implementation of the trend renewal processes for settlement delays. In Section 9, the conclusion follows.

## 2. AGGREGATE TREND RENEWAL PROCESS (ATRP)

Léveillé and Hamel (2017, 2019) have proposed a large class of aggregate claims models that unifies many models encountered in the actuarial literature. Their papers were inspired mainly by Lindqvist et al. (2003), who defined the "trend renewal process" (TRP). This process is mainly used in reliability theory to model recurrent events that incorporate time trends and renewal-type behaviour, such as observed for repairable systems. This large class of stochastic processes will be adapted to the micro-loss reserving framework in this section.

### 2.1. Trend renewal process (TRP)

As mentioned earlier, Lindqvist et al. (2003) rigorously introduced the TRP model in reliability theory. This counting process is a time-transformed renewal process with both the ordinary renewal process and the non-homogeneous Poisson process as special cases. The RP model is related to "perfect repair" conditions, and the NHPP is related to "minimal repair" conditions.

Hence, as in Léveillé and Hamel (2017, 2019) and for the sake of completeness and clarity, it would be appropriate to point out first the formal definition of the TRP given by Lindqvist et al. (2003) before introducing our basic aggregate risk processes.

**Definition 1.** *Let $\lambda(t)$ be a non-negative function defined for $t \geq 0$, satisfying*

$$\Lambda(t) = \int_0^t \lambda(u)\, du < \infty \ , \ t \geq 0 \ \text{ and } \ \Lambda(\infty) = \infty \ .$$

*Furthermore, let $F$ be a positive distribution function (i.e. $F(0) = 0$). Then, the sequence of occurrence times $\{T_k; k \in N^* = \{1, 2, ...\}\}$ is said to generate a $TRP[F, \lambda(.)]$ if the sequence of transformed times $\{\Lambda(T_k); k \in N^*\}$ forms an ordinary renewal process (see Appendix A or Rolski et al., 1999), for which the inter-occurrence times have a common distribution function $F$ and a deterministic (i.e. non-random) trend function $\lambda(t)$.*

**Remark 1.** (1) According to the previous definition, the "events" $T_{k-1} = T_k$ and $\Lambda(T_k) = \Lambda(T_{k-1})$ are impossible since $F(0) = 0$. Then $\Lambda(T_k) - \Lambda(T_{k-1}) > 0$, which implies that $\Lambda$ is a strictly increasing function for $T_{k-1} < T_k$ and then an inverse function $\Lambda^{-1}$ exists.

(2) Two important TRP: a $TRP[F, \lambda]$, where $\lambda$ is a positive constant, is an ordinary renewal process with inter-occurrence time distribution function $F(\lambda x)$, and a $TRP[1 - e^{-x}, \lambda(.)]$ is a non-homogeneous Poisson process with a non-constant intensity function $\lambda(.)$.

(3) If $\{N(t):t\geq 0\}$ is the counting process generated by a TRP$[F,\lambda]$, $f$ is the corresponding density function, $\bar{F}(x)=1-F(x)$ and $\Sigma_{t-}$ is the $\sigma$-algebra of all events just before time $t$, then the conditional intensity function is given by the following expression (see Lindqvist et al. (2003)):

$$\lim_{h\to 0}\frac{P(N(t+h)-N(t)>0|\Sigma_{t-})}{h} = \frac{f\left(\Lambda(t)-\Lambda\left(T_{N(t-)}\right)\right)}{\bar{F}\left(\Lambda(t)-\Lambda\left(T_{N(t-)}\right)\right)}\times \lambda(t) ,$$

where the first factor is the hazard rate function that depends on the transformed time from the last previous failure, and the second one is a factor that depends on the system's age $t$, a factor that could reflect other age-related changes that could occur in the process environment.

### 2.2. Aggregate trend renewal process (ATRP) – Micro model

Let us remind that a dataset, in a loss triangle context, is grouped in a table by the occurrence year of the accident and its development year (age or lag), i.e. the (integer) number of years from the origin year to the (complete) settlement of the claim. The *i*-th accident year is positioned at line *i*, and its *j*-th development year is positioned at column *j*. It is also assumed that all the relevant information is known in the cells of the upper part of a matrix of dimension $t\times t$, including its diagonal, where *t* is the number of accident (and development) years[4] considered.

Very often, the information contained in these cells is incremental net losses. We try to predict them in the cells under the diagonal of this matrix by different techniques. On our part, to calculate (and simulate) these predictions, we will first introduce our ATRP model for each cell, i.e. we consider the following processes:

$$Z_{i,j}(t) = \sum_{k=1}^{N(t)} I(T_k,\xi_k,\zeta_k;t,i,j)\{A_1(T_k+\xi_k+\zeta_k)X_k + A_2(T_k+\xi_k+\zeta_k)Y_k\} , 1\leq i,j\leq t ,$$

where

- $\{T_k;k\in N^*\}$ is a TRP generating the counting process $\{N(t);t\geq 0\}$, such that $T_k$ represents the occurrence time of the *k*-th "accident".
- $\{\xi_k;k\in N^*\}$ is a TRP, independent of the sequence $\{T_k;k\in N^*\}$, such that $\xi_k$ represents the delay taken by the insured from $T_k$ to send the claim to the insurer.
- $\{\zeta_k;k\in N^*\}$ is a TRP, independent of the sequence $\{T_k;k\in N^*\}$ and $\{\xi_k;k\in N^*\}$, such that $\zeta_k$ represents the delay taken by the insurer from $T_k+\xi_k$ to pay the indemnity and the expenses (both payments made at the same time).

---

[4] One year = 365 days

- $\{X_k; k \in N^*\}$ is a sequence of non-negative independent and identically (iid) random variables, independent of sequences $\{T_k; k \in N^*\}$ and $\{\xi_k; k \in N^*\}$, such that $X_k$ represents the deflated amount of the *k*-th claim paid.

- $\{Y_k; k \in N^*\}$ is a sequence of non-negative iid random variables, independent of the sequences $\{T_k; k \in N^*\}$ and $\{\xi_k; k \in N^*\}$, such that $Y_k$ represents the deflated amount of the *k*-th expenses paid.

- The random variables $X_k, Y_k, \zeta_k$ are possibly dependent.

- $A_k(x) = \exp\left\{-\int_t^x \beta_k dv + \int_0^t \alpha_k dv\right\}$, $k = 1, 2$, are the net discount factors at time $t$ corresponding to the payments of the claim and expenses, where $\alpha_k$ and $\beta_k$ are the corresponding constant forces of inflation and (gross) interest.

- $I(T_k, \xi_k, \zeta_k; t, i, j)$ is a product of indicator functions which depend on the random variables $T_k, \xi_k, \zeta_k$ and $t$, and possibly on the *i*-th accident year and the *j*-th development year corresponding to the *k*-th claim.

**Remark 2.** These assumptions are somewhat different from those given in Léveillé and Hamel (2017) since $\{\xi_k; k \in N^*\}$ and $\{\zeta_k; k \in N^*\}$ are assumed to generate TRP's and not only RP's, $\alpha_k$ and $\beta_k$ are constant forces of inflation and interest instead of being both stochastic, and finally the indicator function $I(T_k, \xi_k, \zeta_k; t, i, j)$ depends also of two other quantities, *i* and *j*.

Now, the relevant information to construct our conditional ATRP micro model will be given for $2 \leq i \leq t$, by the following sets:

$$B_i = \{k \in N^* : T_k = t_k, \xi_k = \xi_k, i-1 < t_k + \xi_k \leq i, t < t_k + \xi_k + \zeta_k \leq t+i-1\}, N_{B_i}(t) = n_{B_i},$$

where $B_i$ is related to the set of information corresponding to the claims and expenses of the *i*-th accident year not yet paid in the first *t* development years (strict lower part of the matrix), $n_{B_i}$ is the number of claims in the *i*-th accident year not yet paid and

$$N_{B_i}(t) = \sum_{k=1}^{N(t)} \left( I_{]i-1,i]}(T_k + \xi_k) I_{]t,t+i-1]}(T_k + \xi_k + \zeta_k) \right).$$

Hence, by assuming that $I(T_k, \xi_k, \zeta_k; t, i, j) = I_{]i-1,i]}(T_k + \xi_k) I_{]i+j-2,i+j-1]}(T_k + \xi_k + \zeta_k)$, our conditional ATRP cell predictions will be defined for $2 \leq i \leq t$, $t+2-i \leq j \leq t$, by the following random variables:

$$W_{i,j}(t) = Z_{i,j}(t) | B_i, N_{B_i}(t) = n_{B_i}$$
$$= \sum_{k=1}^{n_{B_i}} I_{]i+j-2, i+j-1]}\left(t_k^{B_i} + \xi_k^{B_i} + \zeta_k^{B_i}\right)\left\{A_1\left(t_k^{B_i} + \xi_k^{B_i} + \zeta_k^{B_i}\right)X_k^{B_i} + A_2\left(t_k^{B_i} + \xi_k^{B_i} + \zeta_k^{B_i}\right)Y_k^{B_i}\right\} | B_i,$$

and our conditional ATRP micro model for the total of predicted payments can be written as follows:

$$W(t) = \sum_{i=2}^{t} \sum_{j=t+2-i}^{t} W_{i,j}(t).$$

**Remark 3.** (1) In the preceding definitions of $W_{i,j}(t)$ and $W(t)$, $X_k, Y_k$ and $\zeta_k$ are the only random variables that we have to model (or calibrate), and the superscript $B_i$ is used to indicate clearly that these random variables and the relevant information are related to the set $B_i$.

(2) If we assume that $I(T_k, \xi_k, \zeta_k; t, i, j) = I_{]0,t]}(T_k + \xi_k) I_{]t, t+i-1]}(T_k + \xi_k + \zeta_k)$, $W(t)$ can also be written as follows:

$$W(t) = \sum_{i=2}^{t} \sum_{k=1}^{n_{B_i}} I_{]t, t+i-1]}\left(t_k^{B_i} + \xi_k^{B_i} + \zeta_k^{B_i}\right)\left\{A_1\left(t_k^{B_i} + \xi_k^{B_i} + \zeta_k^{B_i}\right)X_k^{B_i} + A_2\left(t_k^{B_i} + \xi_k^{B_i} + \zeta_k^{B_i}\right)Y_k^{B_i}\right\} | B_i.$$

Now that we have defined our conditional ATRP micro model as the total of our predictions, usually called the "reserve" in the context of run-off triangles, we will have to build our predictions by considering the expectations of the preceding quantities, along with their standard deviations, coefficients of variation (CV), and their distributions functions. In the next sections, we present the formulas and methods used in this paper to get those quantities for our conditional ATRP micro model.

Although the ATRP governs the event timing in the present implementation, the reserve aggregation formulas derived in Sections 2.2–2.3 remain valid for any alternative parametric arrival and delay processes satisfying the same conditional independence structure.

### 2.3. ATRP micro model - first and second conditional moments

To estimate the expectations of our cell and total predictions according to our conditional ATRP micro model, we have to calculate the following quantities:

$$E\left[W_{i,j}(t)\right], E\left[W(t)\right].$$

Furthermore, it will also be necessary to know their variability, in terms of their variances, by calculating the following quantities:

$$E\left[W_{i,j}^{2}(t)\right],\ E\left[W_{i,j}(t)W_{k,l}(t)\right],\ E\left[W^{2}(t)\right].$$

*2.3.1. ATRP micro model - first conditional moments*

In the following theorem, the first conditional moment of our ATRP process (incremental prediction) is obtained for each cell of the lower triangle and the first conditional moment of the total of these incremental predictions.

**Theorem 1.** *According to the hypotheses and definitions of Sections 2.1 and 2.2, the first moments of $W_{i,j}(t)$ and $W(t)$ are given for $2 \leq i \leq t$, $t+2-i \leq j \leq t$, by the following identities*:

(1) $$E\left[W_{i,j}(t)\right] = \sum_{k=1}^{n_{B_i}} \int_{i+j-2-t_k^{B_i}-\xi_k^{B_i}}^{i+j-1-t_k^{B_i}-\xi_k^{B_i}} \left\{ A_1\left(t_k^{B_i}+\xi_k^{B_i}+v\right) E\left[X_k^{B_i}\Big|\zeta_k^{B_i}=v\right] \right.$$
$$\left. + A_2\left(t_k^{B_i}+\xi_k^{B_i}+v\right) E\left[Y_k^{B_i}\Big|\zeta_k^{B_i}=v\right] \right\} dF_{\zeta_k^{B_i}|B_i}(v),$$

where $$F_{\zeta_k^{B_i}|B_i}(v) = \frac{F_{\zeta_k^{B_i}}(v) - F_{\zeta_k^{B_i}}\left(t-t_k^{B_i}-\xi_k^{B_i}\right)}{F_{\zeta_k^{B_i}}\left(t+i-1-t_k^{B_i}-\xi_k^{B_i}\right) - F_{\zeta_k^{B_i}}\left(t-t_k^{B_i}-\xi_k^{B_i}\right)},\ v \in \left]t-t_k^{B_i}-\xi_k^{B_i}, t+i-1-t_k^{B_i}-\xi_k^{B_i}\right].$$

(2) $$E\left[W(t)\right] = \sum_{i=2}^{t}\sum_{j=t+2-i}^{t} E\left[W_{i,j}(t)\right]$$
$$= \sum_{i=2}^{t}\sum_{k=1}^{n_{B_i}} \int_{t-t_k^{B_i}-\xi_k^{B_i}}^{t+i-1-t_k^{B_i}-\xi_k^{B_i}} \left\{ A_1\left(t_k^{B_i}+\xi_k^{B_i}+v\right) E\left[X_k^{B_i}\Big|\zeta_k^{B_i}=v\right] \right.$$
$$\left. + A_2\left(t_k^{B_i}+\xi_k^{B_i}+v\right) E\left[Y_k^{B_i}\Big|\zeta_k^{B_i}=v\right] \right\} dF_{\zeta_k^{B_i}|B_i}(v).$$

**Proof.** Identity (1) is obtained by taking the conditional expectations of $\zeta_k^{B_i}$ on $B_i$, and identity (2) comes directly from the second definition of $W(t)$. □

*2.3.2. ATRP micro model - second conditional moments*

In the next theorem, the second conditional moment of our ATRP process is obtained for each cell of the lower triangle and the second conditional moment is the total of these incremental predictions.

**Theorem 2.** *According to the hypotheses and definitions of Sections 2.1 and 2.2, the second moments of $W_{i,j}(t)$ and $W(t)$ are given by the following identities*:

(1) $E\left[W_{i,j}^2(t)\right] = \sum_{k=1}^{n_{B_i}} \int_{i+j-2-t_k^{B_i}-\xi_k^{B_i}}^{i+j-1-t_k^{B_i}-\xi_k^{B_i}} \left\{ A_1^2\left(t_k^{B_i}+\xi_k^{B_i}+v\right) E\left[\left(X_k^{B_i}\right)^2 \Big| \zeta_k^{B_i} = v\right] \right.$

$\left. + A_2^2\left(t_k^{B_i}+\xi_k^{B_i}+v\right) E\left[\left(Y_k^{B_i}\right)^2 \Big| \zeta_k^{B_i} = v\right] \right\} dF_{\zeta_k^{B_i}|B_i}(v)$

$+ 2\left\{ \sum_{k=1}^{n_{B_i}-1} \int_{i+j-2-t_k^{B_i}-\xi_k^{B_i}}^{i+j-1-t_k^{B_i}-\xi_k^{B_i}} A_1\left(t_k^{B_i}+\xi_k^{B_i}+v\right) E\left[X_k^{B_i} \Big| \zeta_k^{B_i} = v\right] dF_{\zeta_k^{B_i}|B_i}(v) \right.$

$\times \sum_{r=k+1}^{n_{B_i}} \int_{i+j-2-t_r^{B_i}-\xi_r^{B_i}}^{i+j-1-t_r^{B_i}-\xi_r^{B_i}} A_1\left(t_r^{B_i}+\xi_r^{B_i}+w\right) E\left[X_r^{B_i} \Big| \zeta_r^{B_i} = w\right] dF_{\zeta_r^{B_i}|B_i}(w)$

$+ \sum_{k=1}^{n_{B_i}-1} \int_{i+j-2-t_k^{B_i}-\xi_k^{B_i}}^{i+j-1-t_k^{B_i}-\xi_k^{B_i}} A_2\left(t_k^{B_i}+\xi_k^{B_i}+v\right) E\left[Y_k^{B_i} \Big| \zeta_k^{B_i} = v\right] dF_{\zeta_k^{B_i}|B_i}(v)$

$\left. \times \sum_{r=k+1}^{n_{B_i}} \int_{i+j-2-t_r^{B_i}-\xi_r^{B_i}}^{i+j-1-t_r^{B_i}-\xi_r^{B_i}} A_2\left(t_r^{B_i}+\xi_r^{B_i}+w\right) E\left[Y_r^{B_i} \Big| \zeta_r^{B_i} = w\right] dF_{\zeta_r^{B_i}|B_i}(w) \right\}$

$+ 2\sum_{k=1}^{n_{B_i}-1} \int_{i+j-2-t_k^{B_i}-\xi_k^{B_i}}^{i+j-1-t_k^{B_i}-\xi_k^{B_i}} A_1\left(t_k^{B_i}+\xi_k^{B_i}+v\right) A_2\left(t_k^{B_i}+\xi_k^{B_i}+v\right) E\left[X_k^{B_i} Y_k^{B_i} \Big| \zeta_k^{B_i} = v\right] dF_{\zeta_k^{B_i}|B_i}(v)$

$+ 2\sum_{k=1}^{n_{B_i}-1} \int_{i+j-2-t_k^{B_i}-\xi_k^{B_i}}^{i+j-1-t_k^{B_i}-\xi_k^{B_i}} A_1\left(t_k^{B_i}+\xi_k^{B_i}+v\right) E\left[X_k^{B_i} \Big| \zeta_k^{B_i} = v\right] dF_{\zeta_k^{B_i}|B_i}(v)$

$\times \sum_{r=k+1}^{n_{B_i}} \int_{i+j-2-t_r^{B_i}-\xi_r^{B_i}}^{i+j-1-t_r^{B_i}-\xi_r^{B_i}} A_2\left(t_r^{B_i}+\xi_r^{B_i}+w\right) E\left[Y_r^{B_i} \Big| \zeta_r^{B_i} = w\right] dF_{\zeta_r^{B_i}|B_i}(w)$

$+ 2\sum_{k=1}^{n_{B_i}-1} \int_{i+j-2-t_k^{B_i}-\xi_k^{B_i}}^{i+j-1-t_k^{B_i}-\xi_k^{B_i}} A_2\left(t_k^{B_i}+\xi_k^{B_i}+v\right) E\left[Y_k^{B_i} \Big| \zeta_k^{B_i} = v\right] dF_{\zeta_k^{B_i}|B_i}(v)$

$\times \sum_{r=k+1}^{n_{B_i}} \int_{i+j-2-t_r^{B_i}-\xi_r^{B_i}}^{i+j-1-t_r^{B_i}-\xi_r^{B_i}} A_1\left(t_r^{B_i}+\xi_r^{B_i}+w\right) E\left[X_r^{B_i} \Big| \zeta_r^{B_i} = w\right] dF_{\zeta_r^{B_i}|B_i}(w).$

$$(2)\ E\left[W^2(t)\right] = \sum_{i=2}^{t}\sum_{j=t+2-i}^{t} E\left[W_{i,j}^2(t)\right] + \left\{ \sum_{i=2}^{t}\sum_{j=t+2-i}^{t}\sum_{\substack{l=t+2-i \\ l\neq j}}^{t} E\left[W_{i,j}(t)W_{i,l}(t)\right] \right.$$

$$\left. + \sum_{i=2}^{t}\sum_{j=t+2-i}^{t}\sum_{\substack{k=2 \\ k\neq i}}^{t}\sum_{l=t+2-k}^{t} E\left[W_{i,j}(t)\right]E\left[W_{k,l}(t)\right] \right\},$$

where, for $l \neq j$,

$$E\left[W_{i,j}(t)W_{i,l}(t)\right]$$

$$= \sum_{k=1}^{n_{B_i}-1} \int_{i+j-2-t_k^{B_i}-\xi_k^{B_i}}^{i+j-1-t_k^{B_i}-\xi_k^{B_i}} \left\{ A_1\left(t_k^{B_i}+\xi_k^{B_i}+v\right)E\left[X_k^{B_i}\big|\zeta_k^{B_i}=v\right] + A_2\left(t_k^{B_i}+\xi_k^{B_i}+v\right)E\left[Y_k^{B_i}\big|\zeta_k^{B_i}=v\right] \right\} dF_{\zeta_k^{B_i}|B_i}(v)$$

$$\times \sum_{r=k+1}^{n_{B_i}} \int_{i+l-2-t_r^{B_i}-\xi_r^{B_i}}^{i+l-1-t_r^{B_i}-\xi_r^{B_i}} \left\{ A_1\left(t_r^{B_i}+\xi_r^{B_i}+w\right)E\left[X_r^{B_i}\big|\zeta_r^{B_i}=w\right] + A_2\left(t_r^{B_i}+\xi_r^{B_i}+w\right)E\left[Y_r^{B_i}\big|\zeta_r^{B_i}=w\right] \right\} dF_{\zeta_r^{B_i}|B_i}(w)$$

$$+ \sum_{r=1}^{n_{B_i}-1} \int_{i+l-2-t_r^{B_i}-\xi_r^{B_i}}^{i+l-1-t_r^{B_i}-\xi_r^{B_i}} \left\{ A_1\left(t_r^{B_i}+\xi_r^{B_i}+v\right)E\left[X_r^{B_i}\big|\zeta_r^{B_i}=v\right] + A_2\left(t_r^{B_i}+\xi_r^{B_i}+v\right)E\left[Y_r^{B_i}\big|\zeta_r^{B_i}=v\right] \right\} dF_{\zeta_r^{B_i}|B_i}(v)$$

$$\times \sum_{k=r+1}^{n_{B_i}} \int_{i+j-2-t_k^{B_i}-\xi_k^{B_i}}^{i+j-1-t_k^{B_i}-\xi_k^{B_i}} \left\{ A_1\left(t_k^{B_i}+\xi_k^{B_i}+w\right)E\left[X_k^{B_i}\big|\zeta_k^{B_i}=w\right] + A_2\left(t_k^{B_i}+\xi_k^{B_i}+w\right)E\left[Y_k^{B_i}\big|\zeta_k^{B_i}=w\right] \right\} dF_{\zeta_k^{B_i}|B_i}(w).$$

**Proof.** See Appendix B.

□

## 3. AN ATRP MODEL CALIBRATION – A CASE STUDY: FLORIDA DATASET

In this section, we apply the model defined in Section 2 to the Florida dataset studied in Léveillé and Hamel (2017), drawn from the "First Professionals Insurance Company Inc." to illustrate different coverages for medical malpractice in the USA. Still, this time, we evaluate the conditional expectations, variances, and distributions of the (cell and total) predicted payments. Hence, in Section 3.1, we first describe the dataset used for our study more thoroughly with an analysis of its heterogeneity. In Section 3.2, we first briefly describe the (initial) calibration of the additive model used by the preceding authors and apply our formulas to the chosen dataset. In Section 3.3, we approximate the density function of the total predictions. Finally, in Section 3.4, a short discussion about our calibration with regard to the parameter uncertainty is presented.

### 3.1. The Florida dataset

To illustrate our methodology, a real closed claims database on medical malpractice from the state of Florida in the USA[5] is used to calibrate our model. This Florida database contains detailed information on each claim related to malpractice insurance that has been closed, such as the claim paid, the expense paid, the occurrence times of the accident, the claim and the final payments. As mentioned earlier, we restrict ourselves to the dataset of one of the most important insurers in medical malpractice, more precisely from 2005/01/01 to 2015/01/01, as prior information does not have the mention "Passed Validation" for the record status and then should not be used preferably. Therefore, we will focus on a 10-year historical dataset with enough credible data to calibrate the model considered.

---
[5] https://apps.fldfs.com/PLCR/Search/Home.aspx?Type=External

The whole Florida dataset related to medical malpractice contains nine types of injury, divided into permanent and temporary ones. Permanent injuries include death, grave (quadriplegia, severe brain damage, lifelong care or fatal prognosis), major (paraplegia, blindness, loss of two limbs, brain damage), minor (loss of fingers, loss or damage to organs. Includes non-disabling injuries) and significant (deafness, loss of limb, loss of eye, loss of one kidney or lung). The temporary injuries categories are major (burns, surgical material left, drug side effect, brain damage and recovery delayed), minor (Infections, missed fracture, fall in hospital and Recovery delayed) and slight (lacerations, contusions, minor scars, rash and no recovery delay). The different types of injuries, categories and sub-categories create heterogeneity between individuals and enrich the analysis through the ATRP individual-based model.

An important question arises: What is the degree of heterogeneity among the various types of injuries in this database? Several statistical tools, including Higgins' and Thompson's statistic, are available for our analysis. This statistic is interpreted as the percentage of variability attributed to heterogeneity between groups (injury types) rather than sampling error. Alternatively, an ANOVA test on the means of payments among these groups can also be employed. In this discussion, we will focus on detailing the first method, acknowledging that the second method also provides strong evidence of differences in means across the various injury types. These analyses are conducted using a substantial dataset comprising 63,056 files.

To calculate $I^2$, we first calculate Cochran's $Q$ statistic, which, under the null hypothesis of no heterogeneity, follows a $\chi^2$ distribution with $k$ - 1 degrees of freedom, which are given by

$$Q = \sum_{i=1}^{k} w_i \left( x_i - \frac{\sum_{j=1}^{k} w_j x_j}{\sum_{j=1}^{k} w_j} \right), \quad I^2 = \max\left\{0, \frac{Q-(k-1)}{Q}\right\}.$$

We want to test whether the type of injury significantly impacts claims payments. We use a given type of injury as the experimental group and all other types of injuries as the control group. For the indemnities, we obtain $I^2 = 97.3\%$. According to the range of interpretations given by Higgins et al. (2019), this shows considerable heterogeneity between different types of injuries. Further, we can obtain which group has the most impact on the indemnities paid. The table below describes the different weights and shows that "Permanent: Death" has the most significant impact, with 24.8%.

TABLE 1
Indemnities percentage paid by type of injury

| Injury Type | % |
|---|---|
| Emotional only – Fright, no physical damage. | 5.8 |
| Permanent: Death. | 24.8 |
| Permanent: Grave – quadriplegia, severe brain damage, lifelong care or fatal prognosis. | 4.5 |
| Permanent: Major – Paraplegia, blindness, loss of two limbs, brain damage. | 7.1 |
| Permanent: Minor – Loss of fingers, loss or damage to organs. Includes non-disabling injuries. | 13.7 |
| Permanent: Significant – Deafness, loss of limb, loss of eye, loss of one kidney or lung. | 9.6 |
| Temporary: Major – Burns, surgical material left, drug side effect, brain damage. Recovery delayed. | 11.3 |
| Temporary: Minor – Infections, missed fracture, fall in hospital. Recovery delayed. | 17.3 |
| Temporary: Slight – Lacerations, contusions, minor scars, rash. No delay. | 5.9 |

Similarly, for the expense payments, we obtain $I^2 = 92.4\%$ This means that different types of injuries considerably lead to different expenses. Again, "Permanent: death" also has the most significant impact on the expenses paid.

Therefore, these measures prove considerable heterogeneity in this database, which is also reflected in our selected dataset, and the different types of injuries significantly impact the costs of the claims and expenses paid. As for many other micro models in the actuarial literature, this heterogeneity will be somehow captured in our model by the parameters of our distributions, the possible dependence relation between the claim's payments, expenses payments and delay from the claim to the settlement payment, and by the trend functions (see Lindqvist (2003)). Otherwise, a much more elaborate analysis by subcategories of our database would have to be carried out, which is possible but would represent considerable statistical work.

It is important to clarify that the assumption of claims being settled in a single payment arises as a limitation of the dataset rather than an inherent restriction of the proposed model. The described Florida dataset used in this study provides information on only a single aggregated payment to settle the claim, as it is a closed-claim database. However, the model itself is not constrained to this assumption. If such information were available, it could be extended to account for multiple payments by incorporating additional data, such as individual payment amounts and corresponding timestamps. This flexibility ensures the methodology remains adaptable and applicable to more complex reserving scenarios in real-world contexts, where claims are often settled through multiple payments.

### 3.2. Example of an ATRP model calibration

To justify the distributional assumptions in our ATRP model, we emphasize that the selection process was driven by thorough exploratory data analysis and rigorous model-fitting procedures. Multiple candidate distributions were evaluated using goodness-of-fit criteria, including the Akaike Information Criterion (AIC) and Bayesian Information Criterion (BIC), to ensure the best possible representation of the data. For instance, the Generalized Gamma distribution was chosen for modeling settlement delays due to its superior empirical fit, as determined through these selection methods. The calibration process, including detailed statistical considerations and parameter estimation techniques, follows the methodology outlined in Léveillé and Hamel (2017), where truncated distributions and dependence structures were systematically addressed. Furthermore, the calibration methodology accounts for the unique characteristics of the Florida dataset, leveraging its historical depth to refine parameter estimation and ensure robustness. We refer the interested reader to Section 3 of Léveillé and Hamel (2017) for a comprehensive discussion on the choice and justification of the model's distributions. Given these foundations, we proceed under the following assumptions:

- $(X_k, Y_k, T_k, \xi_k, \zeta_k)$ are the retained data from the closed-claims database such that $5519 \text{ days } (15.13 \text{ years}) \leq T_k + \xi_k + \zeta_k \leq 9171 \text{ days } (25.13 \text{ years})$, where $T_k$ is the k-th occurrence time of the accident satisfying the preceding inequalities, $\xi_k$ and $\zeta_k$ are the corresponding delays, $X_k$ and $Y_k$ are the corresponding deflated payments. Also, $T_0 = 0$ corresponds to the date $1989/11/22$.

- $\{X_k \equiv \tilde{X}_k (1 + 365\, \zeta_k)^{\kappa_1}; k \in N^*\}$ is a sequence of non-negative iid random variables independent of the $T_k$ and $\xi_k$, $\tilde{X}_k$ is independent of $\zeta_k$ and the calibrated $\kappa_1 = 0.29504$. The relationship between $\tilde{X}_k$ and $\zeta_k$ is chosen as a functional form for simplicity and interpretability in the calibration process. By introducing this structure, we can observe a natural link between indemnity amounts and delays, which aligns with empirical patterns observed in practice. While alternative dependence structures could be adopted, this specific formulation provides flexibility in capturing various types of dependencies. The parameter $\kappa_1$ plays a critical role in this context: when $\kappa_1 = 0$, independence is achieved; positive values of $\kappa_1$ indicate positive dependence, while negative values capture negative dependence. This functional approach thus enables a wide range of dependency scenarios between claim amounts and delays while maintaining a manageable complexity for calibration.

- $\{Y_k \equiv \tilde{Y}_k (1 + 365\, \zeta_k)^{\kappa_2}; k \in N^*\}$ is a sequence of non-negative iid random variables independent of the $T_k$ and $\xi_k$, $\tilde{Y}_k$ is independent of $\tilde{X}_k$ and $\zeta_k$, and the calibrated $\kappa_2 = 1.23178$.

- The density function of $\zeta_k$ is given by the following generalized gamma:

$$f_\zeta(x) = \frac{1}{x\Gamma(a)} b \left(\frac{x}{c}\right)^{ab} \exp\left\{-\left(\frac{x}{c}\right)^b\right\},$$

where,

TABLE 2

Parameters calibration of the density function of $\zeta_k$

| $a$ | $b$ | $c$ |
|---|---|---|
| 3.33246873 | 0.67977335 | 0.3645056 |

- The dependence between $X_k$ and $Y_k$ being only through $\zeta_k$, their calibrated distributions functions are assumed to be given by the following expressions:

$$F_{X_k|\zeta_k}(x|t) = p_X + (1-p_X)\left\{\sum_{i=1}^{2} c_{X,i} \int_0^x \frac{1}{v}\frac{1}{\sqrt{2\pi}\sigma_{X,i}} \exp\left[-\frac{(\ln v - \mu_{X,i} - \kappa_1 \ln(1+365\,t))^2}{2\sigma_{X,i}^2}\right] dv\right\},$$

where,

TABLE 3

Parameters calibration of the density function of $X_k|\zeta_k$

| $p_X$ | $c_{X,1}$ | $c_{X,2}$ | $\mu_{X,1}$ | $\mu_{X,2}$ | $\sigma_{X,1}$ | $\sigma_{X,2}$ |
|---|---|---|---|---|---|---|
| 0.5605836 | 0.7193306 | 0.2806694 | 8.590078 | 9.603317 | 1.316284 | 0.2598194 |

and,

$$F_{Y_k|\zeta_k}(x|t) = p_Y + (1-p_Y)\left\{\sum_{i=1}^{2} c_{Y,i} \int_0^x \frac{1}{v}\frac{1}{\sqrt{2\pi}\sigma_{Y,i}} \exp\left[-\frac{(\ln v - \mu_{Y,i} - \kappa_2 \ln(1+365\,t))^2}{2\sigma_{Y,i}^2}\right] dv\right\},$$

where,

TABLE 4

Parameters calibration of the distribution function of $Y_k|\zeta_k$

| $p_Y$ | $c_{Y,1}$ | $c_{Y,2}$ | $\mu_{Y,1}$ | $\mu_{Y,2}$ | $\sigma_{Y,1}$ | $\sigma_{Y,2}$ |
|---|---|---|---|---|---|---|
| 0.1683231 | 0.3142661 | 0.6857334 | -0.05958437 | 0.9696933 | 1.1458589 | 0.7298423 |

- For the payments of claims, the calibrated force of inflation is $\alpha_1 = 0.045692$, and the force of interest is assumed to be $\beta_1 = 0.06$.

- For the payments of expenses, the calibrated force of inflation is $\alpha_2 = 0.041744$, and the force of interest is assumed to be $\beta_2 = 0.06$.

**Remark 4.** (1) The forces of inflation are estimated on the retained dataset and reapplied on the deflated amounts of indemnities and expenses to get our predictions.

(2) In practice, the payments of claims and expenses are not necessarily made simultaneously, and these payments can be spread out. Their related discount rates are not necessarily known and equal, and the inflation and discount rates can change from one year to another. Consequently, to avoid too much complexity in our example, we will assume in the next section that the inflation and discount rates remain constant for all the years related to our triangle. Furthermore, we will assume that claims and expenses are paid simultaneously.

**Application of the previous calibration**

As a first application of the previous calibration of our conditional ATRP model, we will use the formulas presented in Section 2.3 to get the predictions of the incremental payments ($E[W_{ij}(t)]$) in Table 5, their standard deviations ($\sigma[W_{ij}(t)]$) in Table 6, the total of these predictions ($E[W(t)]$) and its standard deviation ($\sigma[W(t)]$). Furthermore, the numbers presented in each cell $(i, j)$ of the lower triangle are the inflated payments in the development year $j$, discounted at the beginning of the first year of prediction for the accident year $i$.

Our predictions for 10 accident years and 10 development years will be estimated. These incremental payments are in US dollars (rounded to the dollar). Our formulas are evaluated by the Cuhre method in **R** Package "cubature."

TABLE 5

$E[W_{ij}(t)]$, with $\alpha_1 = 0.045692$, $\alpha_2 = 0.041744$ and $\beta_1 = \beta_2 = 0.06$

| i\j | 1 | 2 | 3 | 4 | 5 | 6 | 7 | 8 | 9 | 10 |
|---|---|---|---|---|---|---|---|---|---|---|
| 1 | | | | | | | | | | |
| 2 | | | | | | | | | | 1,369,474 |
| 3 | | | | | | | | | 2,025,207 | 1,268,932 |
| 4 | | | | | | | | 1,064,960 | 667,006 | 417,858 |
| 5 | | | | | | | 2,855,697 | 1,793,982 | 1,123,575 | 703,693 |
| 6 | | | | | | 4,754,728 | 3,006,133 | 1,886,665 | 1,181,310 | 739,947 |
| 7 | | | | | 8,972,332 | 5,774,152 | 3,651,174 | 2,291,453 | 1,434,621 | 898,490 |
| 8 | | | | 8,713,198 | 5,816,327 | 3,737,492 | 2,361,603 | 1,481,606 | 927,462 | 580,846 |
| 9 | | | 3,560,946 | 2,533,705 | 1,671,490 | 1,068,480 | 673,540 | 422,154 | 264,207 | 165,502 |
| 10 | | 151,440 | 158,579 | 117,474 | 78,683 | 50,626 | 32,002 | 20,077 | 12,565 | 7,866 |

## TABLE 6
$\sigma[W_{ij}(t)]$, with $\alpha_1 = 0.045692$, $\alpha_2 = 0.041744$ and $\beta_1 = \beta_2 = 0.06$

| i\j | 1 | 2 | 3 | 4 | 5 | 6 | 7 | 8 | 9 | 10 |
|---|---|---|---|---|---|---|---|---|---|---|
| 1 | | | | | | | | | | |
| 2 | | | | | | | | | | 1,197,731 |
| 3 | | | | | | | | | 1,533,174 | 1,260,141 |
| 4 | | | | | | | | 1,118,022 | 912,036 | 736,328 |
| 5 | | | | | | | 1,819,806 | 1,482,596 | 1,194,925 | 957,815 |
| 6 | | | | | | 2,330,247 | 1,903,200 | 1,534,974 | 1,229,967 | 982,604 |
| 7 | | | | | 3,153,304 | 2,601,816 | 2,108,123 | 1,692,056 | 1,351,814 | 1,077,937 |
| 8 | | | | 3,044,936 | 2,569,081 | 2,103,321 | 1,696,247 | 1,357,636 | 1,082,799 | 862,541 |
| 9 | | | 1,897,953 | 1,665,460 | 1,386,949 | 1,127,473 | 905,684 | 723,312 | 576,223 | 458,761 |
| 10 | | 359,761 | 396,656 | 355,626 | 298,329 | 243,194 | 195,550 | 156,203 | 124,411 | 99,006 |

The mean, standard deviation, and coefficient of variation (CV) obtained (from our formulas) for the total predicted payments are respectively $82,489,260, $9,048,247 and 11%, which means that the mean remains representative of the total predicted payments, even if the standard deviation may be considered significant (in million dollars) for an actuary depending of his risk aversion. The total of the predicted claims payments, which is $57,918,295, represents approximately 70.2% of total predicted payments.

**Remark 5.** (1) For most of the prediction cells, it can be easily checked that the CV is relatively high in comparison with the CV of the total predicted payments. This may be partly explained by a relatively small number of payments jointly with relatively large spreads between the payments in each cell and partly by the calibration method used on the historical dataset.

(2) If we had assumed that $X_k, Y_k, \zeta_k$ were stochastically independent (by letting $\kappa_1 = \kappa_2 = 0$ and recalibrating by the procedure described previously), then the calculations of the mean, standard deviation, and coefficient of variation (CV) of the total predicted payments would have been $64,974,100, $8,369,742, and 12.9% (see Table 14 further below).

(3) If we had assumed that the two inflation rates were zero (again implying a recalibration by the procedure described previously), then the calculations of the mean, standard deviation and coefficient of variation (CV) of the total predicted payments would have been respectively $60,632,633, $6,529,588 and 10.8% (see further below, Table 12).

## 3.3. Approximation of the density function of the total predictions

It is not generally possible to get an exact expression of the density function (and distribution) of such a complex stochastic process. Still, many statistical tools are at our disposal, especially if we have enough data, to get at least a representation of this density function, such as given by Gaussian Kernel density estimation. However, a parametric expression would be more helpful to get a deeper insight into the risk process.

Hence, we will use our previous example and make 100,000 simulations of this particular conditional ATRP model to approximate its density function by a normal mixture, i.e. we assume that its density is given by

$$f(x) = \lambda_1 \frac{1}{\sqrt{2\pi}\sigma_1} \exp\left(-\frac{(x-\mu_1)^2}{2\sigma_1^2}\right) + \lambda_2 \frac{1}{\sqrt{2\pi}\sigma_2} \exp\left(-\frac{(x-\mu_2)^2}{2\sigma_2^2}\right), \lambda_1 \geq 0, \lambda_2 \geq 0, \lambda_1 + \lambda_2 = 1.$$

Using the Expectation-Maximization (EM) algorithm and the **R** package 'mixdist', we thus get the following table and graphic of this approximate density function.

TABLE 7
Normal mixture parameters fit related to Example 3.2

| $i$ | $\lambda_i$ | $\mu_i$ | $\sigma_i$ |
|---|---|---|---|
| 1 | 0.8679651 | 80,900,516 | 7,142,919 |
| 2 | 0.1428373 | 92,969,347 | 12,443,256 |

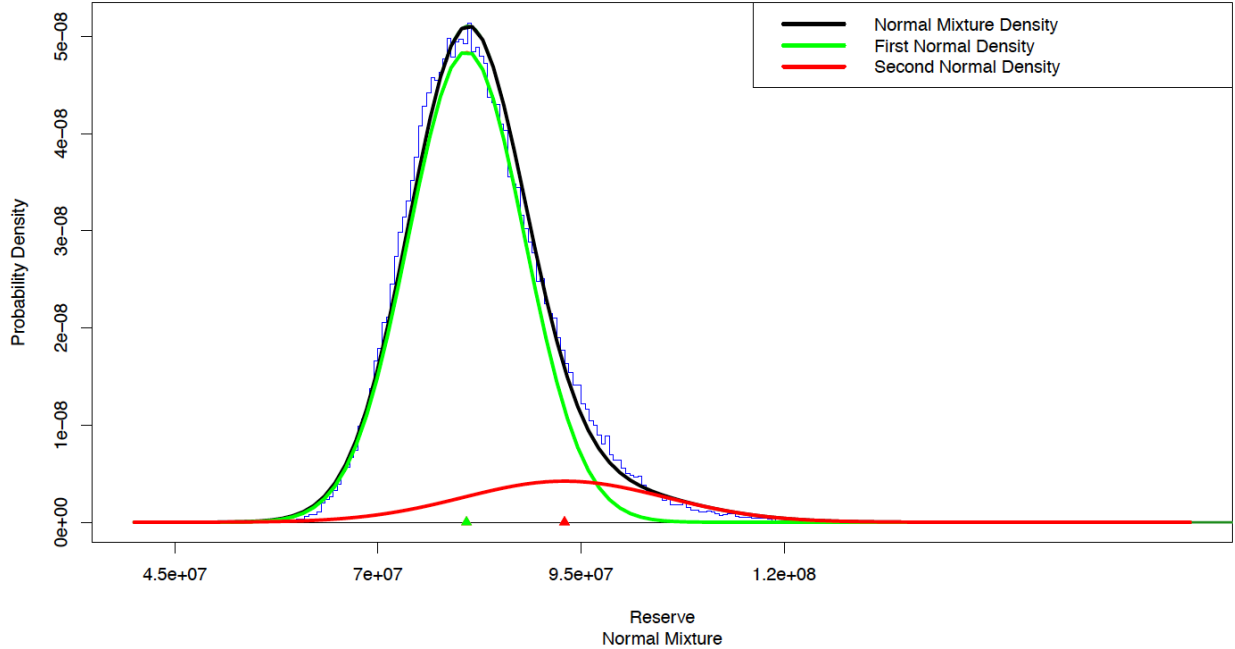

FIGURE 1. Normal mixture density fit

As we can see, by using 100,000 simulations of our conditional ATRP model, the simulated mean and standard deviation given in Table 8 are relatively close to the corresponding quantities calculated in Section 3.2. The density function also presents a very accurate fit on this empirical distribution, which is also corroborated by their absolute and (absolute) relative differences with respect to the simulated values, in the following table.

TABLE 8
Absolute errors and absolute relative errors for the normal mixture fit

| ... | Mean | St. dev. | VaR60 | VaR80 | VaR95 | TVaR60 | TVaR80 | TVaR95 |
|---|---|---|---|---|---|---|---|---|
| Simul. | 82,494,972 | 9,023,393 | 83,631,891 | 89,077,709 | 98,221,641 | 90,999,883 | 95,863,290 | 105,279,882 |
| N. mixt. | 82,605,972 | 9,130,472 | 83,857,617 | 88,952,942 | 98,868,608 | 91,131,210 | 96,053,594 | 106,344,151 |
| Abs. diff. | 111,000 | 107,079 | 225,726 | 124,767 | 646,967 | 131,327 | 190,304 | 1,064,269 |
| A.Rel.diff. | 1.3% | 1.2% | 0.3% | 0.1% | 0.7% | 0.1% | 0.2% | 1% |

This Normal mixture fit could be improved by increasing the number of simulations and (or) Normals. It is also possible to use other statistical fitting methods in this context, like Gamma mixture or GAM fitting. Furthermore, having a good approximation of our conditional ATRP model allows us to look more closely at a vast range of risk measures other than the usual VaR, TVaR and loss ratio, such as stop-loss premiums or distortion risk measures. The exact fitting method could also be applied to our conditional ATRP model with other inflation and interest rates or dependence structures; ordering these distributions between them will be much easier if we have good approximations.

### 3.4. Parameters uncertainty

Parameter uncertainty, related to the calibration of our conditional ATRP model, usually has to be considered in order to evaluate the risk incurred by the model in practice. Using an algorithm to analyze the uncertainty of the parameters (see Appendix C), we generate 100,000 scenarios and thus get the following table of the mean, standard deviation, VaR and TVaR values and their increased percentage from the corresponding simulated values obtained in Table 8.

TABLE 9
Mean, standard deviation, VaR, TVaR with parameters uncertainty

| … | Mean | St. dev. | VaR60 | VaR80 | VaR95 | TVaR60 | TVaR80 | TVaR95 |
|---|---|---|---|---|---|---|---|---|
| Values with p.u. | 85,751,918 | 14,216,533 | 86,935,529 | 95,579,133 | 110,722,861 | 98,916,891 | 106,974,814 | 123,462,342 |
| % incr. | 3.8% | 36.5% | 3.8% | 6.8% | 11,3% | 8.0% | 10.4% | 14.7% |

These results explain why so many insurers, presenting higher levels of risk aversion, incorporate parameter uncertainty, which gives more conservative results but may be less competitive since the premiums have to follow the risk calculations. Other procedures, such as Markov Chain Monte Carlo methods (see Rafferty et al. (1996)), could also be used to reassure the insurers about the different results obtained and make the best business decisions.

## 4. CHANGES IN OUR CONDITIONAL ATRP ASSUMPTIONS

In this section, we are mainly interested in analyzing the "sensitivity" of our conditional ATRP model to different changes in our previous assumptions. We will focus our study on the total reserve, and it would be interesting to study these changes in each prediction cell changes on each prediction cells. First, we will examine the impact of different inflation and discount rates on the reserve. Secondly, we will look at the dependence relation between the payments, expenses, and delay and see how this affects the reserve. Lastly, we will examine a certain change in the settlement delay. All this will be done by 100,000 simulations (without parameter uncertainty), and we will compare most of these tables against each other.

### 4.1 Changes in inflation and discount rates

This section analyzes the impact of different inflation and discount rates on the reserve's mean, standard deviation and some selected Var and TVaR values. For changes only on discount rates, the (calibrated) inflation rates and the discount rates are applied to the calibrated deflated claims and expenses, such as in subsection 3.2. For changes only in inflation rates, we have to recalibrate $\kappa_1, \kappa_2$ and the conditional distributions of $X_k|\zeta_k$ and $Y_k|\zeta_k$. Afterward, the inflation rates and the (fixed) discount rates are applied to the calibrated deflated claims and expenses.

### 4.1.1. Discount rates $\beta_1 = \beta_2$ at 0%, 2%, 4%, 6%.

In this subsection, we keep all the hypotheses and calibrated values obtained in Section 3.2, except we now assume that $\beta_1 = \beta_2$ and successively equal to 0%, 2%, 4%, 6%. Hereafter, in Table 10, the results were obtained from 100,000 simulations.

TABLE 10

$\beta_1 = \beta_2$ at 0%, 2%, 4%, 6%

| $\beta_1 = \beta_2$ | Mean | St. dev. | VaR60 | VaR80 | VaR95 | TVaR60 | TVaR80 | TVaR95 |
|---|---|---|---|---|---|---|---|---|
| 0% | 91,911,613 | 10,141,678 | 93,177,194 | 99,295,740 | 109,505,918 | 101,461,086 | 106,919,815 | 117,506,577 |
| 2% | 88,576,063 | 9,732,329 | 89,802,757 | 95,668,782 | 105,474,029 | 97,744,303 | 102,984,836 | 113,138,947 |
| 4% | 85,442,036 | 9,361,053 | 86,618,217 | 92,270,746 | 101,715,523 | 94,263,515 | 99,306,538 | 109,072,743 |
| 6% | 82,494,024 | 9,023,393 | 83,631,891 | 89,077,709 | 98,221,641 | 90,999,883 | 95,863,290 | 105,279,882 |

As expected, all the selected quantities decrease as the discount rates increase. The CVs are nearly the same for these discount rates (11.0%), and then the means remain representative of the total predicted payments at relatively the same level regardless of the discount rates. For our discount rates going from 0% to 6%, all the selected quantities (means, st. dev., …) are each reduced by approximately 10% and 11%.

### 4.1.2. Discount rates $\beta_1 \neq \beta_2$ at (2%,4%), (4%,2%), (4%,6%), (6%,4%).

In this subsection, we keep all the hypotheses and calibrated values obtained in Section 3.2, except we now assume that $\beta_1 \neq \beta_2$ and equal to (2%,4%), (4%,2%), (4%,6%), (6%,4%). Hereafter, in Table 11, the results were obtained from 100,000 simulations.

TABLE 11

$\beta_1 \neq \beta_2$ at (2%,4%), (4%,2%), (4%,6%), (6%,4%).

| $(\beta_1, \beta_2)$ | Mean | St. dev. | VaR60 | VaR80 | VaR95 | TVaR60 | TVaR80 | TVaR95 |
|---|---|---|---|---|---|---|---|---|
| (2%,4%) | 87,524,053 | 9,707,992 | 88,739,425 | 94,581,373 | 104,401,017 | 96,667,548 | 101,900,994 | 112,050,526 |
| (4%,2%) | 86,494,045 | 9,385,422 | 87,682,770 | 93,345,859 | 102,784,574 | 95,340,266 | 100,390,423 | 110,160,407 |
| (4%,6%) | 84,456,659 | 9,340,138 | 85,627,116 | 91,259,014 | 100,710,150 | 93,256,940 | 98,293,949 | 108,056,747 |
| (6%,4%) | 83,479,401 | 9,044,305 | 84,622,083 | 90,092,022 | 99,224,102 | 92,006,383 | 96,875,688 | 106,295,576 |

Again, we see that all the selected quantities are decreasing from top to bottom of the table. The CV are nearly the same for these discount rates (11.1%, 10.9%, 11.1%, 10.8%) and the means remain representative of the predicted payments regardless of the discount rates. For our discount rates going from the top to the bottom of the table, the mean and the standard deviation of the total predicted payments are reduced approximately by 4.6% and 6.8%, and for the other quantities the reduction varies between 4.6% and 5.1%. If you compare the two last lines of Table 11 to the last line of Table 10, you observe that all the numbers of these two lines in Table 11 are greater than those corresponding in Table 10, this because one of the discount rates has been reduced by 2% thus increasing the reserve. We also note that this increase is more important for the case $(\beta_1, \beta_2) = (4\%, 6\%)$ since the indemnities payments represent approximatively 70% of the total predicted payments. The first two lines of Table 11 are motivated by the same arguments.

### 4.1.3. Inflation rates $\alpha_1 = \alpha_2$ at 0%, 2%, 4%, 6%.

In this subsection, we assume that $\alpha_1 = \alpha_2$ and equal to 0%, 2%, 4%, 6%. For each choice of the inflation rates, we recalibrate $\kappa_1, \kappa_2$, the conditional distributions of $X_k | \zeta_k$ and $Y_k | \zeta_k$, and we keep all the other calibrated distributions and discount values ($\beta_1 = \beta_2 = 6\%$) presented in Section 3.2. Hereafter, in Table 12, the results were obtained from 100,000 simulations.

TABLE 12
$\alpha_1 = \alpha_2$ at 0%, 2%, 4%, 6%

| $\alpha_1 = \alpha_2$ | Mean | St. dev. | VaR60 | VaR80 | VaR95 | TVaR60 | TVaR80 | TVaR95 |
|---|---|---|---|---|---|---|---|---|
| 0% | 60,602,837 | 6,564,998 | 61,446,569 | 65,423,480 | 72,142,759 | 66,816,387 | 70,358,976 | 77,193,972 |
| 2% | 72,195,589 | 6,631,351 | 73,129,636 | 77,101,761 | 83,698,009 | 78,467,979 | 81,965,802 | 88,618,050 |
| 4% | 93,552,844 | 7,239,315 | 94,745,932 | 99,138,467 | 106,053,152 | 100,485,887 | 104,179,155 | 110,945,521 |
| 6% | 131,787,616 | 9,264,533 | 133,537,958 | 139,122,838 | 147,564,875 | 140687,296 | 145,222,771 | 153,241,379 |

We see that all the selected quantities are increasing as the inflation rates increase. The CV is 10.8%, 9.2%, 7.7%, and 7.0% respectively, and then the representativity of the means increases when the inflation rates increase. For our inflation rates going from 0% to 6%, the mean and the standard deviation of the total predicted payments are increased by 117.5% and 41.1%. For the other quantities, the increases vary between 98.5% and 117.3%. If you compare the third line in Table 12, where $\alpha_1 = \alpha_2 = 4\%$, to the last line of Table 10, where $\alpha_1 = 0.045692$ and $\alpha_2 = 0.041744$, then you observe a decrease of the mean by 11.8% and an increase of the standard deviation by 24.6%, and for the other quantities, this reduction varies between 5.1% and 11.7%. So, keeping $\beta_1 = \beta_2 = 6\%$, even relatively small reductions in inflation rates, approximately 0.57% for expenses and (mainly) 0.17% for claims can have an important impact on reserve reduction.

*4.1.4. Inflation rates $\alpha_1 \neq \alpha_2$ at (2%,4%), (4%,2%), (4%,6%), (6%,4%)*

In this subsection, we assume that $\alpha_1 \neq \alpha_2$ and equal to (2%,4%), (4%,2%), (4%,6%), (6%,4%). For each choice of inflation rates, we must recalibrate as described previously in Subsection 4.1.3, and we keep $\beta_1 = \beta_2 = 6\%$. Hereafter, in Table 13, the results were obtained from 100,000 simulations.

TABLE 13

$\alpha_1 \neq \alpha_2$ (2%,4%), (4%,2%), (4%,6%), (6%,4)

| $(\alpha_1, \alpha_2)$ | Mean | St. dev. | VaR60 | VaR80 | VaR95 | TVaR60 | TVaR80 | TVaR95 |
|---|---|---|---|---|---|---|---|---|
| (2%,4%) | 84,639,723 | 6,134,641 | 85,753,154 | 89,409,519 | 95,120,556 | 90,511,807 | 93,566,136 | 99,083,849 |
| (4%,2%) | 83,969,265 | 8,311,345 | 85,096,164 | 90,123,564 | 98,419,424 | 91,832,795 | 96,255,071 | 104,751,958 |
| (4%,6%) | 122,798,164 | 8,352,085 | 124,479,995 | 129,535,820 | 137,048,754 | 130,861,998 | 134,893,379 | 141,694,721 |
| (6%,4%) | 105,346,897 | 8,914,469 | 106,695,004 | 112,005,701 | 120,810,126 | 113,802,989 | 118,447,758 | 127,264,800 |

The CV is respectively 7.2%, 9.9%, 6.8%, and 8.4%, and then the means remain representative of the total predicted payments. If you compare the case $(\alpha_1, \alpha_2) = (6\%, 4\%)$ in Table 13 to the case $\alpha_1 = \alpha_2 = 4\%$ in Table 12, then you observe that increasing $\alpha_1$ by 2% increases the mean by 31.1% and the standard deviation by 57.3%. Now, if we consider the case $(\alpha_1, \alpha_2) = (4\%, 6\%)$ in Table 13 to the case $\alpha_1 = \alpha_2 = 4\%$ in Table 12, we see that increasing $\alpha_2$ by 2% increases the mean by 52.8% and the standard deviation by 47.4%. These differences can be explained by the fact that the recalibration of the case $(\alpha_1, \alpha_2) = (6\%, 4\%)$, caused by the increase of 2% of $\alpha_2$, has given "more importance" to the expenses compared to the claims than the recalibration for the case $(\alpha_1, \alpha_2) = (4\%, 6\%)$. A similar analysis could be done if you compare the cases $(\alpha_1, \alpha_2) = (2\%, 4\%)$ and $(\alpha_1, \alpha_2) = (4\%, 2\%)$ to the case $\alpha_1 = \alpha_2 = 4\%$.

## 4.2 Changes in the dependence relation between $X_k, Y_k$ and $\zeta_k$

In this section, we analyze the impact of the dependence relation between $X_k, Y_k$ and $\zeta_k$ on the reserve's mean, standard deviation and some selected VaR and TVaR values. We will first look at the case where $X_k, Y_k$ and $\zeta_k$ are stochastically independent, which means that $\kappa_1 = \kappa_2 = 0$ in the calibration method described in Section 3.2, and thereafter, we will examine the case where $X_k$ and $Y_k$ are stochastically dependent but independent of $\zeta_k$.

### 4.2.1. If $X_k, Y_k$ and $\zeta_k$ are independent

In this subsection, we now assume that $\kappa_1 = \kappa_2 = 0$, we recalibrate the distributions of $X_k$ and $Y_k$, and we keep all the other hypotheses and calibrated values obtained in Section 3.2, except that we look at different values of $\beta_1 = \beta_2$. Hereafter, in Table 14, the results were obtained from 100,000 simulations.

TABLE 14
$\beta_1 = \beta_2$ , $X_k, Y_k$ and $\zeta_k$ are independent

| $\beta_1 = \beta_2$ | Mean | St. dev. | VaR60 | VaR80 | VaR95 | TVaR60 | TVaR80 | TVaR95 |
|---|---|---|---|---|---|---|---|---|
| 0% | 71,764,040 | 9,252,223 | 72,782,424 | 78,300,295 | 87,783,936 | 80,338,035 | 85,358,610 | 95,320,831 |
| 2% | 69,366,306 | 8,931,635 | 70,353,011 | 75,666,738 | 84,853,410 | 77,638,631 | 82,480,611 | 92,082,505 |
| 4% | 67,106,418 | 8,638,516 | 68,065,088 | 73,189,849 | 82,089,645 | 75,102,589 | 79,782,020 | 89,057,419 |
| 6% | 64,974,100 | 8,369,742 | 65,909,482 | 70,868,331 | 79,462,143 | 72,716,751 | 77,247,626 | 86,227,964 |

We see that all the selected quantities decreased as the discount rates increase, as in Table 10. The CVs are all the same (after rounding to one decimal) for these discount rates (12.9%), and then the means remain representative of the total predicted payments. For our discount rates going from 0% to 6%, all the selected quantities related to the total predictions are each reduced by approximately 9.5%. If you compare Table 10 to Table 14, for the corresponding discount rates of 0%, 2%, 4%, and 6%, the percentages of decrease of the means and standard deviations vary respectively between 21.2% and 21.9% and between 7.2%, and 8.7%, and for the other quantities vary approximatively between 18.1% and 21.8%. This shows that the positive dependence between $X_k, Y_k$ and $\zeta_k$ in Table 10 has a great impact on the reserve.

### 4.2.2. If $X_k, Y_k$ are dependent but independent of $\zeta_k$

In this subsection, we now assume that $X_k$ and $Y_k$ are independent of $\zeta_k$, with the (recalibrated) mixtures of lognormals as in Section 3.2, and are related by a Franck copula, i.e.

$$C(u,v) = -\frac{1}{\theta} \ln \left\{ 1 + \frac{(e^{-\theta u} - 1)(e^{-\theta v} - 1)}{e^{-\theta} - 1} \right\} , \; u,v \in [0,1], \; \theta \neq 0$$

Using the "fitCopula" R-package, with the "itau" method of estimation, we get $\theta = 1.413523$. Hereafter, in Table 15, the results were obtained from 100,000 simulations.

TABLE 15

$\beta_1 = \beta_2$, $X_k, Y_k$ are dependent and independent of $\zeta_k$

| $\beta_1 = \beta_2$ | Mean | St. dev. | VaR60 | VaR80 | VaR95 | TVaR60 | TVaR80 | TVaR95 |
|---|---|---|---|---|---|---|---|---|
| 0% | 71,252,450 | 10,704,331 | 72,193,903 | 77,441,143 | 86,031,113 | 79,810,431 | 84,987,255 | 97,398,836 |
| 2% | 68,872,973 | 10,315,222 | 69,784,987 | 74,846,209 | 83,148,222 | 77,127,835 | 82,119,914 | 94,088,190 |
| 4% | 66,630,348 | 9,960,097 | 67,519,371 | 72,413,419 | 80,430,366 | 74,607,676 | 79,431,926 | 90,994,606 |
| 6% | 64,514,365 | 9,635,113 | 65,368,796 | 70,104,549 | 77,849,450 | 72,236,830 | 76,908,302 | 88,097,255 |

Again, we see that all the selected quantities are decreasing as the discount rates increase. The CVs are nearly the same for these discount rates (15.0%). For our increasing discount rates, all the selected quantities related to the total prediction are each reduced by approximately 9.5%. Suppose you compare Table 15 to Table 10. In that case, this new dependence assumption has a substantial increasing impact of approximately 28% for the mean but a slight decrease between 5.2% and 6.3% for the standard deviation. We also note that the results of Table 15 are slightly less than those of Table 14, just by adding dependence only between $X_k, Y_k$ by the Franck Copula.

### 4.3. Change in the settlement delay process

This section examines the impact of a change in the settlement delay. This can be motivated by a staff turnover at a specific point in time, which could cause a certain slowdown in the payment operations or, on the contrary, which will speed up the payment process and reduce the reserve. To illustrate this possible change, we will change a parameter in our generalized gamma distribution of the settlement delay (for the claims unsettled in the first ten years). We will assume that the settlement delay is the calibrated generalized gamma of Section 3.2, with the only difference being that the parameter c is divided by 2, i.e.

$$f_\zeta(x) = \frac{0.67977335}{x\Gamma(3.33246873)} \left(\frac{x}{0.1822528}\right)^{(3.33246873)(0.67977335)} \exp\left\{-\left(\frac{x}{0.1822528}\right)^{0.67977335}\right\}.$$

By this choice, the mean delay settlement is approximatively reduced by half with respect to the original calibrated distribution of $\zeta_k$, which will reduce the reserve relative to the initial calibration. Table 16 presents the results obtained from 100,000 simulations.

TABLE 16

$\beta_1 = \beta_2$, generalized gamma with parameter c divided by 2

| $\beta_1 = \beta_2$ | Mean | St. dev. | VaR60 | VaR80 | VaR95 | TVaR60 | TVaR80 | TVaR95 |
|---|---|---|---|---|---|---|---|---|
| 0% | 82,102,255 | 9,237,208 | 83,330,768 | 88,895,916 | 98,358,960 | 90,871,358 | 95,852,806 | 105,422,120 |
| 2% | 80,366,486 | 9,023,541 | 81,562,874 | 87,006,142 | 96,251,100 | 88,933,936 | 93,800,056 | 103,139,431 |
| 4% | 78,700,139 | 8,823,939 | 79,871,194 | 85,197,664 | 94,230,027 | 87,078,764 | 91,836,891 | 100,964,313 |
| 6% | 77,099,315 | 8,637,073 | 78,245,164 | 83,463,850 | 92,299,675 | 85,300,829 | 89,958,086 | 98,890,030 |

Similarly, we observe that the mean, standard deviation and risk measures decrease as the discount rates increase. The coefficient of variation (CV) is nearly the same for these discount rates, and around 11.2%, which confirms that the means remain equally representative. When we increase the discount rate from 0% to 6%, all the selected quantities are reduced by approximately 6.1%, which was expected by the decrease in settlement delays. Again, if we compare Table 16 to Table 10 with increasing discount rates, the corresponding means and standard deviations decrease from 6.5% to 10.7% and from 6.1% to 10.4%.

## 4.4 Incorporating Claim Type as a Covariate

*4.4.1. Introduction*

This section extends the Aggregate Trend Renewal Process (ATRP) to include claim type as a covariate, addressing the limitations of earlier models that assumed homogeneity across claims. These assumptions often lead to inflated reserve estimates and reduced predictive accuracy. By integrating claim type as a covariate, the model accounts for heterogeneity in settlement delays and payment distributions, refining calibration and reliability. This extension significantly enhances the accuracy of reserve predictions while maintaining the robustness of the ATRP model.

*4.4.2. Methodology*

Incorporating claim type into the ATRP involves recalibrating the settlement delay distribution separately for each of the nine injury categories, ranging from temporary injuries to permanent disabilities. Additionally, a single model was calibrated across all nine classes, incorporating claim type as a covariate for both payment amounts and expenses. This approach leverages claim-specific data to enhance the granularity of the model, allowing it to capture variability in settlement patterns, indemnities, and expenses across different claim types, ultimately improving the precision of reserve estimates.

To model this heterogeneity, we introduce a covariate, representing the claim type into the probability density functions for settlement delay and payment size. Specifically:

$$F_{Y_k|\zeta_k}(x|t) = p_Y + (1-p_Y)\left\{\sum_{i=1}^{2} c_{Y,i} \int_0^x \frac{1}{v} \frac{1}{\sqrt{2\pi}\sigma_{Y,i}} \exp\left[-\frac{(\ln v - \mu_{Y,i} - \kappa_2 \ln(1+365\,t) - \Phi_Y N_Y)^2}{2\sigma_{Y,i}^2}\right] dv\right\}$$

where $\Phi_Y$ is a vector of parameters of size 9 and $N_Y$ is a zero-one vector indicating the class of size 9. The settlement delay is modeled as a function of claim type to account for differences in the average settlement time between injury categories.

$$F_{X_k|\zeta_k}(x|t) = p_X + (1-p_X)\left\{\sum_{i=1}^{2} c_{X,i} \int_0^x \frac{1}{v}\frac{1}{\sqrt{2\pi}\sigma_{X,i}} \exp\left[-\frac{\left(\ln v - \mu_{X,i} - \kappa_1 \ln(1+365\,t) - \Phi_X N_X\right)^2}{2\sigma_{X,i}^2}\right]dv\right\}$$

where $\Phi_X$ is a vector of parameters of size 9 and $N_X$ is a zero-one vector indicating the class of size 9. The payment size distribution incorporates the claim type covariate to reflect the variability in indemnities and expenses among claim types.

The segmented dataset enables the estimation of claim-specific parameters, capturing differences in average settlement time and indemnities across categories. The model recalibration incorporates these parameters to improve the precision of predictions.

*4.4.3. Results and Analysis*

Table 17 presents the recalibrated reserve estimates, incorporating claim type as a covariate. These updated estimates reflect the model's ability to account for claim type heterogeneity and suggest improved predictive performance compared to the original specification. The formal validation of these predictions against the actual RBNS reserves from the data is presented later, in Section 4.5, where we demonstrate the enhanced accuracy achieved by the proposed approach.

TABLE 17
Incorporation of the Claim Type as a covariate

| $\beta_1 = \beta_2$ | Mean | St. dev. | VaR60 | VaR80 | VaR95 | TVaR60 | TVaR80 | TVaR95 |
|---|---|---|---|---|---|---|---|---|
| 0% | 74,677,943 | 16,477,822 | 76,051,974 | 86,536,439 | 104,730,559 | 90,341,779 | 99,839,273 | 118,110,233 |
| 2% | 71,988,600 | 15,893,257 | 73,306,844 | 83,442,106 | 100,947,987 | 87,097,822 | 96,262,925 | 113,892,354 |
| 4% | 69,460,900 | 15,348,060 | 70,733,385 | 80,524,250 | 97,443,283 | 84,052,369 | 92,907,291 | 109,941,138 |
| 6% | 67,082,456 | 14,838,648 | 68,319,870 | 77,754,023 | 94,160,842 | 81,189,984 | 89,754,565 | 106,233,737 |

The recalibrated model demonstrates a reduction in mean reserves across all discount rates. For example, the mean reserve at a 0% discount rate decreases from approximately 91.9 million (original model) to 74.7 million, primarily due to the improved differentiation of claims by type. This reduction mitigates the inflationary effects of assuming homogeneity.

While mean reserve estimates decrease, tail metrics, such as TVaR95, slightly increase. For instance, TVaR95 rises from 117.5 million to 118.1 million at a 0% discount rate. This increase reflects the model's enhanced ability to capture extreme outcomes, ensuring a more conservative risk estimate under adverse scenarios. Furthermore, the recalibrated standard deviations are higher than those in the original model, indicating an improved representation of localized variability.

*4.4.4. Implications and Extensions*

The integration of claim type as a covariate highlights the significance of granular-level predictors in reserve modeling. This refinement reduces reserve estimates, improves tail risk modeling, and enhances variability representation. The results underscore the importance of accounting for claim-specific heterogeneity in loss reserving, as it increases predictive accuracy and provides a clearer understanding of underlying risks.

Future research could extend this framework on a dataset that contains additional predictors, such as geographic or policyholder-specific factors, further enriching the model's applicability. Additionally, integrating economic conditions and external factors could provide valuable insights into claim behaviour and reserve requirements, ensuring a robust and comprehensive actuarial framework.

## 4.5 Model Validation and Sensitivity Analysis

Accurate reserve estimation is a cornerstone of effective risk management and solvency assessment in insurance. The Aggregate Trend Renewal Process (ATRP) model offers significant advantages over traditional methods due to its flexibility in integrating multiple covariates and assumptions. However, its robustness and reliability must be rigorously validated against actual reserves to ensure practical applicability. This section presents reserve estimates as averages, allowing for a more stable, consistent and interpretable comparison across different modeling assumptions and sensitivity analyses.

This section comprehensively validates the ATRP model by comparing its average reserve estimates to the true reserve derived from actual data. In addition, a sensitivity analysis explores the impact of key assumptions—such as inflation, dependency structures, and settlement delay patterns—on reserve estimates. These analyses address critical questions about the ATRP model's predictive accuracy, adaptability under varying conditions, and relevance for practical applications. The results provide evidence of the model's strengths and offer insights into the implications of different modeling assumptions on reserve adequacy.

*4.5.1 Validation Against True Reserve*

The predictive accuracy of the Aggregate Trend Renewal Process (ATRP) model was assessed by comparing reserve estimates against the true reserve of $71,005,064 derived from actual data. The models compared include the Mack model, the ATRP without covariates, and the ATRP incorporating claim type as a covariate. The results are summarized in Table 18.

TABLE 18

Reserve Comparison and Accuracy Metrics with 0% discount rate

| Model | Reserve Estimate | Coefficient of Variation (%) | MAPE (%) |
|---|---|---|---|
| Mack Model | 67,795,372.36 | 20.0 | 5.65 |
| ATRP Without Covariates | 91,911,613 | 11.0 | 16.30 |
| ATRP With Covariates | 74,677,943 | 22.0 | 3.83 |

The results in Table 18 reveal a clear accuracy–stability trade-off among the three models. In terms of point accuracy, the ATRP model with covariates performs best, achieving the lowest MAPE of 3.83%, followed by the Mack model (MAPE = 5.65%), while the ATRP model without covariates exhibits substantially poorer accuracy (MAPE = 16.30%) and a pronounced overestimation of the reserve. With respect to estimation stability, as measured by the coefficient of variation, the Mack model yields the lowest variability (20.0%), closely followed by the ATRP with covariates (22.0%), whereas the ATRP without covariates again performs worst in this dimension (CV = 11.0%, reflecting model misspecification rather than true stability). Overall, incorporating covariates into the ATRP framework substantially improves estimation accuracy while retaining a level of uncertainty comparable to the Mack model, thereby offering a favourable balance between predictive performance and reserve stability.

It is important to note that the Mack model estimate presented in Table 18 corresponds specifically to the RBNS reserve. In this analysis, the input triangle provided to the Mack model was constructed exclusively from incremental payments associated with reported but not yet settled (RBNS) claims as of the valuation date, excluding incurred-but-not-reported (IBNR) liabilities by design. The Mack model produces an estimate of the total reserve for the specific liability represented by the input triangle (see Wüthrich and Merz (2008)). Hence, applying the Mack model to the RBNS triangle yields a reserve that is directly comparable to the RBNS reserves estimated by the ATRP model under the same data scope. This ensures methodological consistency and allows for a meaningful comparison of predictive performance across models within the same liability subset.

*4.5.2 Sensitivity Analysis*

The sensitivity analysis investigates how different assumptions and model configurations influence reserve estimates, providing a deeper understanding of the ATRP model's robustness and predictive accuracy. Table 19 presents reserve estimates, which are the mean of the distribution under various scenarios and their Mean Absolute Percentage Errors (MAPEs) relative to the true reserve value of $71,005,064. These scenarios focus on the impact of assumptions such as inflation, dependency structures, settlement delays, and the inclusion of covariates, allowing a direct comparison of model performance. To enable a meaningful and fair comparison with the

Mack model—commonly used in practice and typically applied without discounting—we work under the assumption of a 0% discount rate.

TABLE 19
Sensitivity Analysis Results (MAPE at 0% Discount)

| Table Reference | Assumption/Configuration | Avg. Reserve Estimate[6] | MAPE (%) |
|---|---|---|---|
| Table 10 | Baseline | 91,911,613 | 29.43 |
| Table 18 | Mack Model | 67,795,372.36 | 5.65 |
| [7]Table 12 | No Inflation ($\alpha_1 = 0, \alpha_2 = 0$) | 67,154,900 | 5.42 |
| Table 15 | Dependent Assumptions | 71,252,450 | 0.35 |
| Table 16 | Reduced Settlement Delays | 82,102,255 | 15.61 |
| Table 17 | Claim Type as Covariate | 74,677,943 | 3.79 |

The baseline model (Table 10), calibrated using historical data without advanced refinements, produces a reserve estimate of $91,911,613, exceeding the true reserve by 29.4%. This overestimation highlights the shortcomings of simplified models that ignore critical features such as inflation, interdependencies among claim components, settlement delays, and claim-type heterogeneity. Such omissions can lead to systematic biases, underscoring the necessity of incorporating these real-world complexities to achieve reliable and accurate reserve estimates.

Excluding inflation assumptions entirely and assuming a 0% interest rate (Table 12) results in a reserve estimate of $67,154,900, underestimating the true reserve by 5.42%. This deviation highlights the pivotal role of inflation in reserve modeling, as it directly affects the magnitude of claims and expenses. Failing to account for inflation leads to significant underestimation, showcasing the necessity of incorporating realistic inflationary trends to ensure accurate predictions.

Incorporating dependencies between indemnities and expenses while assuming independence from settlement delays (Table 15) produces a reserve estimate of $71,252,450, nearly identical to the true reserve (MAPE = 0.3%). This remarkable improvement in accuracy emphasizes the importance of modeling interdependencies among claim variables, reflecting the interconnected nature of real-world claims processes. Such dependency structures enhance the model's ability to capture nuanced dynamics, improving its reliability for decision-making.

Adjusting the settlement delay function to reflect shorter settlement times (Table 16) decreases the reserve estimate compared to Table 10 (the baseline), bringing it down to $82,102,255. This

---

6: Here, the reserve estimate is equal to the mean of the reserves distribution to ensure a consistent comparison with the Mack Model.
7: This result is not directly displayed in the original Table 12. Instead, it is derived by adjusting the discount factor to 0% to ensure a consistent comparison with other scenarios in Table 19. All other assumptions from Table 12 remain unchanged.

represents a 15.6% deviation from the true reserve. The sensitivity of reserve estimates to settlement delay assumptions underscores the importance of precise calibration based on empirical data. Even slight variations in settlement timing can significantly impact reserve projections, highlighting the need for assumptions that closely align with observed claim behaviours.

Finally, including claim type as a covariate (Table 17) results in a reserve estimate of $74,677,943, with a deviation of just 3.79%. By accounting for heterogeneity in claim characteristics, this model configuration reduces overestimation observed in the baseline model while enhancing the granularity and predictive power of the reserve estimates. This result illustrates the value of integrating claim-level predictors, which improve accuracy and provide actionable insights for tailoring reserve allocation strategies.

*4.5.3 Discussion and Implications*

The findings from the sensitivity analysis demonstrate the flexibility and robustness of the ATRP model in adapting to diverse assumptions and configurations. These results highlight the model's ability to capture reserve dynamics under varying conditions, offering critical insights for actuarial practice. The baseline configuration provides a starting point for calibration but is limited by its lack of refined assumptions, as evidenced by the significant overestimation of reserves. The model aligns with the true reserve by incorporating realistic factors such as inflation and dependency structures, as shown in Table 19.

The analysis underscores the ATRP model's practical value for insurers and actuaries. For instance, the strong alignment of reserves under dependent assumptions reflects the model's capacity to capture real-world interactions between key variables, supporting better decision-making in capital allocation and risk management. Furthermore, integrating claim type as a covariate enhances the model's predictive accuracy and enables a more granular understanding of claim behaviour, facilitating improved risk segmentation and resource allocation.

These results also emphasize the importance of sensitivity analyses in actuarial modeling. By systematically testing key assumptions, the proposed framework provides insurers with a rigorous tool to evaluate reserve adequacy under a range of scenarios, in alignment with regulatory and solvency requirements. While alternative modeling assumptions and specifications could naturally be considered, the sensitivity of reserve estimates to settlement delay assumptions highlights the critical need for accurate empirical calibration of delay distributions. At the same time, the pronounced impact of inflation underscores its fundamental role in capturing future liabilities and reinforces the necessity of explicitly modeling financial dynamics in long-tailed lines of business.

It is also worth clarifying an important point regarding the interaction between inflation assumptions and claim-level heterogeneity. As shown in the sensitivity analysis, incorporating inflation without simultaneously accounting for claim type heterogeneity results in an overestimation of reserves, as inflation is applied uniformly across all claims, regardless of their settlement dynamics or severity patterns. This phenomenon is particularly evident in the baseline configuration, where the inflation adjustment magnifies reserve estimates beyond observed liabilities. However, once the claim type is included as a covariate (Table 17), the inflation adjustment is applied more appropriately to each claim category, yielding reserve estimates that align more closely with the actual RBNS reserve and achieving a markedly lower MAPE. This illustrates that inflation and heterogeneity should be modeled jointly to avoid biased projections, as failing to account for heterogeneity can distort the impact of inflation. Our framework thus allows actuaries to disentangle and correctly attribute the effects of these factors, demonstrating its practical utility and methodological soundness.

In conclusion, the proposed unified micro-level modeling framework provides a powerful and flexible approach for reserve estimation, balancing predictive accuracy with practical adaptability. By explicitly incorporating real-world complexities such as dependence, operational delays, inflation, and discounting, the framework is well suited to address the challenges of modern insurance practice. While the current implementation relies on a specific stochastic structure, alternative modeling assumptions and specifications could also be considered within the same unified architecture, without altering the core reserving logic.

Future research could further enhance the framework by exploring additional covariates, alternative dependence structures, and other process assumptions, thereby broadening its applicability across a wider range of actuarial contexts.

## 5. INSIGHTS TO UNEARNED PREMIUM RISK, IBNR, and TREND ANALYSIS

### 5.1. Introduction

The IBNR and UPR components developed in this section are embedded within the same unified micro-level reserving framework introduced earlier. Although the future claim trajectories are generated using the ATRP for internal consistency with the RBNS modeling layer, the IBNR proportions and UPR estimates are anchored in empirical calibration on observed data. The ATRP is therefore used solely as a projection mechanism rather than as the defining element of the IBNR methodology. The resulting IBNR estimates are thus not purely synthetic development factors but represent data-driven reserve components evaluated at a defined valuation date.

As such, this section addresses two critical actuarial challenges: unearned premium risk and Incurred But Not Reported (IBNR) claims. These elements are pivotal for actuaries in estimating the reserves required to cover future risks and quantifying the uncertainty surrounding unpaid claims. By leveraging the ATRP model, we simulate claims for future periods to evaluate reserves and assess premium risk.

Incorporating trends is an essential enhancement to the model. Due to the absence of a complete historical dataset (caused by the sale of the underlying insurance company), incorporating trends allows us to reflect real-world dynamics and understand the potential impacts on reserve estimates. This approach ensures that the ATRP model accommodates variations in claim activity over time, offering a more comprehensive and practical framework for reserve modeling.

### 5.2. Trend Analysis and Its Implications

*5.2.1. Observations from Premium Trends*

Table 20 provides a detailed overview of the direct premiums written by FPIC in Florida over time and their corresponding market share. The data reveal a clear upward trajectory in premiums from 2005, peaking in 2011 with $126,798,631 in direct premiums and a market share of 22.8%. However, following the sale of the company to "Doctors Company" in 2011, there was a marked decline in premium volume as business lines were gradually transferred, culminating in the cessation of operations in 2016.

TABLE 20
Direct premiums written by FPIC in Florida and their corresponding market share over the years.

| Year | Direct Premiums Written (USD) | Market Share in Florida (%) |
|---|---|---|
| 2016 | 0 | 0 |
| 2015 | 0 | 0 |
| 2014 | 1,783,341 | 0.3 |
| 2013 | 98,677,086 | 19.1 |
| 2012 | 106,798,631 | 19.6 |
| 2011 | 122,757,236 | 22.8 |
| 2010 | 116,184,713 | 20.8 |
| 2009 | 139,231,343 | 23.5 |
| 2008 | 158,409,971 | 23.9 |
| 2007 | 192,813,921 | 24.8 |
| 2006 | 215,690,159 | 25.4 |
| 2005 | 210,174,357 | 24.2 |

The evolution of premiums underscores the dynamic nature of insurance markets, which are influenced by operational, regulatory, and market-specific factors. A static modeling approach would fail to capture these temporal variations, highlighting the necessity of models capable of accommodating evolving trends. Within the proposed unified micro-level framework, temporal dynamics are incorporated through a flexible intensity specification calibrated to historical premium experience. While the current implementation relies on a specific stochastic formulation for tractability and coherence with the reserving structure, alternative modeling assumptions and intensity specifications could equally be considered within the same framework. This flexibility allows the model to generate realistic projections aligned with observed historical patterns, while preserving robustness and practical applicability.

*5.2.2. Calibrated Temporal Intensity Function*

Figure 2 depicts the estimated intensity function over time, serving as a proxy for the frequency of claim events. We calibrated the trend function below using the rigorous approach outlined by Léveillé and Hamel (2017), leveraging the actual data from the upper triangle to ensure a precise and data-driven calibration. This methodology highlights the model's capability to directly derive and incorporate trend information from real-world datasets rather than relying on assumptions or external inputs. By grounding the calibration in observed data, the approach ensures that the estimated trend function accurately reflects historical claim behaviours and aligns closely with empirical evidence.

The intensity function demonstrates a steep increase from 1990, reaching its peak around 2000, followed by a gradual decline through 2016. This temporal pattern closely mirrors the premium data presented in Table 20, reinforcing the connection between premium volume and claim intensity.

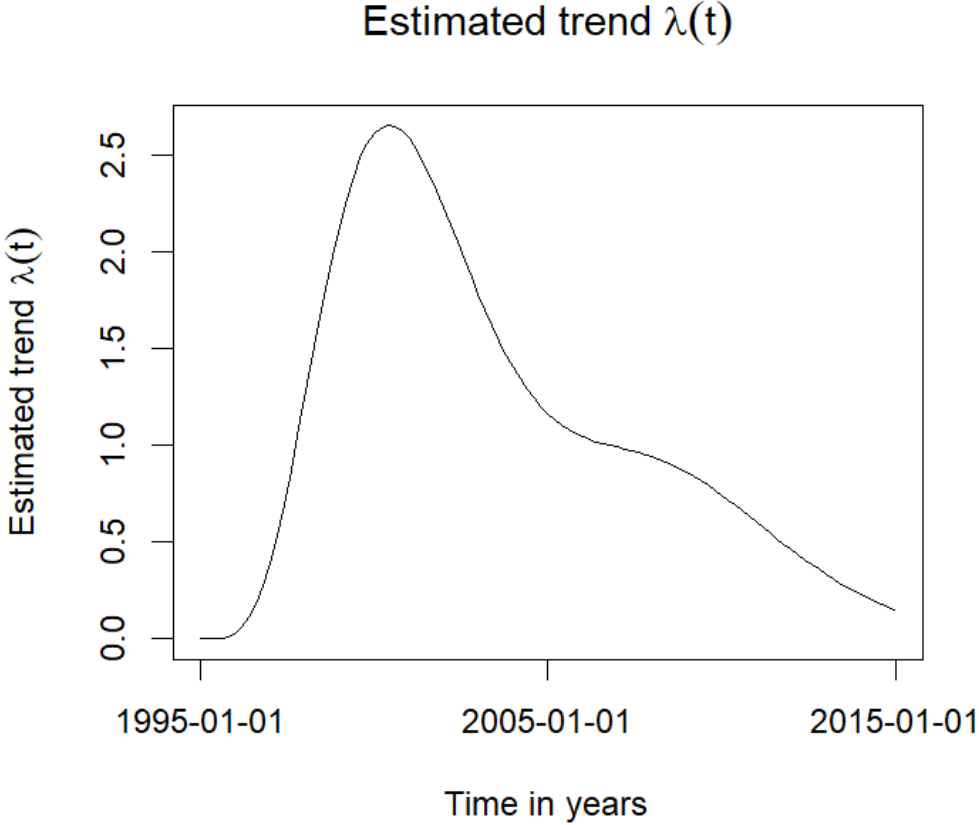

FIGURE 2. Estimated intensity function over time, illustrating the frequency of claim events. The intensity peaks around 2000, followed by a gradual decline until 2016, reflecting market conditions and claim activity changes.

The alignment between the intensity function and premium trends highlights the importance of incorporating temporal effects into reserve models. The peak intensity around 2000 likely reflects heightened claim activity driven by market or policyholder behaviour. Similarly, the decline in intensity after 2005 corresponds to the tapering of premium volumes, suggesting that claim activity is closely tied to market share and business operations. These observations validate the necessity of trend incorporation within our framework.

## 5.3. Methodology

The company was sold, so we can no longer access premium data for subsequent years. Consequently, in this section, we used an arbitrary trend for the following year's numerical analysis to demonstrate the model's practical application. While the trend calibrated earlier (see Figure 2) was derived from real historical data, the arbitrary trend employed here is an illustrative example. This trend, chosen for its simplicity, represents a monotonically increasing function over time that converges asymptotically to infinity. This setup showcases the model's capability to simulate claims for future periods, even when specific data—such as premium volume—are unavailable. For instance, had the premium information been accessible, the model could have been directly utilized to calculate unearned premium reserves, offering valuable insights into the associated risk. The model's flexibility allows insurers to adapt to varying data constraints and still derive actionable predictions. By applying this arbitrary trend, we illustrate how the model can be operationalized to assess and manage future liabilities. We emphasize its robustness and applicability in different practical scenarios, including estimating claims or reserve needs when premium data is missing.

The ATRP model simulated future claims under various trends and interest rate scenarios. Key metrics evaluated in this analysis include:

- Occurrence Claims: The total claims incurred.

- Claims Made: The claims reported.

- Tail Coverage: The difference between occurrence claims and claims made, reflecting IBNR claims.

For a full definition of the Occurrence Claims, Claims Made, and Tail Coverage, we invite the interested reader to review section 2 of Léveillé and Hamel (2017). In this section, we consider the following (arbitrary) trend function:

$$\Lambda(t) = t^{\gamma}$$

where $\gamma > 0$. The model was tested under three interest rate scenarios (0%, 2%, and 6%), enabling the evaluation of reserves using metrics such as mean, Value-at-Risk (VaR), and Tail Value-at-Risk (TVaR). To capture temporal variations, the trend function was adapted based on calibrated intensity functions, following the work of Léveillé and Hamel (2017). The flexibility of the ATRP model allows it to incorporate temporal trends that reflect historical claim intensity and premium dynamics.

## 5.4. Results and Analysis

Table 21 summarizes the simulated outcomes for the most relevant trends, with a fixed interest rate of 6%, focusing on occurrence claims, claims made, and tail coverage. We simulated the following random variables: occurrence, tail-coverage and claims-made processes, respectively, as follows:

$$Z_{occ}(t) = \sum_{k=1}^{N(t)} 1_{\{T_k \leq t\}} \left\{ A_1(T_k + \xi_k + \zeta_k) X_k + A_2(T_k + \xi_k + \zeta_k) Y_k \right\},$$

$$Z_{tc}(t) = \sum_{k=1}^{N(t)} 1_{\{T_k + \xi_k > t\}} \left\{ A_1(T_k + \xi_k + \zeta_k) X_k + A_2(T_k + \xi_k + \zeta_k) Y_k \right\}$$

and

$$Z_{cm}(t) = \sum_{k=1}^{N(t)} 1_{\{T_k + \xi_k \leq t\}} \left\{ A_1(T_k + \xi_k + \zeta_k) X_k + A_2(T_k + \xi_k + \zeta_k) Y_k \right\},$$

over a one-year period where $1_{\{A\}}$ takes the value 1 if the condition $A$ is satisfied and 0 otherwise.

TABLE 21

Simulated Costs, Value-at-Risk, and Tail Value-at-Risk under varying trends, for a one-year period

| Trend | Interest Rate | Metric | Mean | St. Dev. | VaR95 | TVaR95 |
|---|---|---|---|---|---|---|
| $\gamma = 0.5$ | 6% | Occurrence | 1,429,582 | 917,307 | 2,351,293 | 2,909,118 |
| $\gamma = 0.5$ | 6% | Claims Made | 497,779 | 517,550 | 1,246,986 | 2,001,501 |
| $\gamma = 0.5$ | 6% | Tail Coverage (IBNR) | 931,803 | 720,916 | 1,241,163 | 2,032,207 |
| $\gamma = 1.0$ | 6% | Occurrence | 27,030,162 | 6,604,270 | 39,080,355 | 43,205,962 |
| $\gamma = 1.0$ | 6% | Claims Made | 7,339,859 | 2,619,191 | 14,410,532 | 20,015,009 |
| $\gamma = 1.0$ | 6% | Tail Coverage (IBNR) | 19,690,303 | 4,999,211 | 25,764,312 | 29,876,590 |
| $\gamma = 1.5$ | 6% | Occurrence | 515,480,349 | 104,097,597 | 705,441,220 | 776,515,324 |
| $\gamma = 1.5$ | 6% | Claims Made | 115,709,645 | 27,661,354 | 186,002,328 | 215,067,265 |
| $\gamma = 1.5$ | 6% | Tail Coverage (IBNR) | 399,770,704 | 85,764,290 | 523,085,987 | 562,523,429 |

*5.4.1. Impact of Trends and Interest Rates*

The results demonstrate that higher trends substantially increase the simulated costs and tail metrics. For example, the mean occurrence cost rises from \$1.4M with $\gamma = 0.5$ to \$515.5M when $\gamma = 1.5$ under a 6% interest rate. This underscores the critical role of trend assumptions in determining reserve adequacy.

Additionally, lower interest rates result in higher cost estimates for both occurrence claims and claims made, reflecting the discounting effect on future liabilities. VaR and TVaR metrics further reveal the sensitivity of extreme outcomes to trends. For instance, the TVaR95 for occurrence claims increases from \$2.9M when $\gamma = 0.5$ to \$776.5M when $\gamma = 1.5$, emphasizing the elevated risk under high-trend scenarios.

*5.4.2. Tail Coverage (IBNR)*

Including IBNR reserves is crucial for assessing the completeness and robustness of reserve models, particularly for long-tail lines such as medical malpractice. Both RBNS (Reported But Not Settled) and IBNR claims play pivotal roles in medical malpractice insurance. While RBNS claims reflect liabilities already reported to the insurer, IBNR claims represent potential liabilities for incidents that have occurred but have yet to be reported. In many instances, the significance of IBNR claims can vary depending on the reporting behaviour and settlement lag typical to the line of business. Due to the protracted legal and reporting processes, IBNR reserves can constitute a substantial portion of total reserves for medical malpractice. Therefore, a thorough analysis of tail risk metrics, particularly those related to IBNR claims, provides valuable insights into reserve adequacy and risk management.

The results presented in Table 21 illustrate the sensitivity of tail coverage (reflecting IBNR reserves) to different trend assumptions and interest rates. Tail coverage expands significantly as trends increase, driven by the compounding effects of rising claims frequency and severity. For example, under a high-trend scenario ($\gamma = 1.5$) with a 6% interest rate, the mean tail coverage reaches \$399.8M, compared to \$931.8K under a low trend scenario ($\gamma = 0.5$). This stark contrast underscores the importance of modeling trends accurately, as they directly influence reserve projections and risk capital requirements.

Moreover, the tail risk metrics, Value-at-Risk (VaR) and Tail Value-at-Risk (TVaR), further highlight the model's ability to quantify extreme outcomes associated with IBNR claims. For instance, the TVaR for tail coverage rises from \$2.03M under $\gamma = 0.5$ to \$562.5M under $\gamma = 1.5$, reflecting the amplified risk in high-trend scenarios. These findings emphasize the need for insurers to consider IBNR reserves carefully when planning for extreme loss events and managing long-term liabilities.

While this study primarily focuses on RBNS claims, extending our framework to IBNR reserves demonstrates its flexibility and relevance for broader applications. The methodology leverages a trend renewal process (TRP) to capture the claim arrival process, which is fundamental for estimating IBNR reserves. The model's ability to simulate claim occurrences and arrivals provides a dynamic approach to assessing unearned premium risk and projecting future liabilities, particularly when historical premium information is unavailable due to data limitations. In this context, we acknowledge the absence of premium data for future projections due to the sale of the company; however, the model's functionality allows insurers to incorporate such data, if available, to enhance IBNR reserve estimates further.

Finally, the comparative significance of IBNR and RBNS reserves in medical malpractice warrants discussion. While RBNS reserves are often more immediately impactful due to their direct connection to reported claims, the inclusion of IBNR reserves offers a more holistic view of an insurer's liabilities. In scenarios where reporting delays are extended, as is common in medical malpractice, IBNR reserves can become even more critical. The sensitivity analysis presented in this section underscores the importance of integrating IBNR considerations into our framework to improve predictive accuracy and support more informed decision-making in reserve management.

*5.4.3 Justification for Trend Incorporation and Implications for Reserve Modeling*

Integrating trends into the ATRP model is not arbitrary but firmly grounded in empirical evidence and theoretical advancements. The ATRP model generalizes the non-homogeneous Poisson process to incorporate temporal variations in claim activity by calibrating trends such as those observed in the intensity function and premium data. This approach ensures that the model captures real-world fluctuations in claims frequency and severity, highlighting the adequacy of trend representation in reserve models. The calibration process, informed by historical data, allows the model to account for dynamic market conditions, enhancing its capacity to provide realistic and accurate reserve estimates reflective of the actual risk environment.

Incorporating trends within our framework offers significant benefits for reserve modeling and actuarial practice. For instance, the model aligns reserve estimates closely with observed data by capturing the rise-and-fall patterns in premiums and claims. This capability is critical for addressing challenges in modeling reported but not settled (RBNS) claims, as it allows for a more precise representation of liabilities over time. The inclusion of trends also enables the model to reflect the long-term effects of external factors such as economic cycles, regulatory reforms, or shifts in market conditions, ensuring its relevance in contemporary actuarial contexts.

The ATRP model advances traditional approaches, such as the non-homogeneous Poisson process, by incorporating more nuanced trend behaviours and their implications. Unlike static models, our framework allows trends to evolve over time and interact with external covariates, such as claim types, inflation rates, and regulatory shifts. This flexibility provides a deeper understanding of the drivers behind reserve estimates, enabling actuaries to identify and quantify the factors influencing claims frequency, severity, and settlement delays. For example, the calibrated trend in Figure 2 demonstrates how the intensity of claim events mirrors fluctuations in premium volumes, underscoring the strong connection between market activity and reserve requirements.

Additionally, trend incorporation plays a vital role in showcasing the model's application to RBNS reserves. By leveraging trend data, the ATRP model bridges the gap between historical observations and future projections, facilitating reserve estimates that account for both known claims and evolving risk profiles. This capability is further validated through sensitivity analyses, which demonstrate the impact of trend assumptions on reserve adequacy and solvency, as illustrated in Table 21.

The implications of trend incorporation extend beyond reserve estimation to broader actuarial applications, such as solvency assessments, capital allocation, and pricing strategies. By integrating trends into its framework, the ATRP model provides insurers with actionable insights essential for maintaining reserve adequacy and solvency in a rapidly changing insurance landscape. Moreover, its ability to adapt to diverse scenarios, including those with significant temporal variations, solidifies the ATRP model's position as a valuable and robust tool in actuarial science. In conclusion, the observations from Table 21 and Figure 2 highlight the critical importance of trend analysis in reserve modeling. By effectively incorporating trends, the ATRP model ensures its utility as a reliable and versatile framework for managing insurance reserves in complex and dynamic environments.

## 6. Modeling Future Liabilities from IBNR and Unearned Premiums

### 6.1 Motivation and Scope

Incurred But Not Reported (IBNR) reserves represent an insurer's obligation for claims that have occurred but have not yet been reported within a specified time horizon. Often overlooked in conventional reserve analyses, accurately estimating the IBNR proportion is crucial, especially for long-tailed lines such as medical malpractice, where claims frequently emerge with significant delays and severity. Unlike RBNS (Reported But Not Settled) claims, IBNR liabilities pertain directly to events already incurred but remain hidden due to reporting delays.

In this section, we focus specifically on a one-year projection horizon. To robustly estimate future liabilities from unreported claims, we introduce an innovative approach based on the Aggregate Trend Renewal Process (ATRP). This methodology simulates both the frequency and severity of claims occurring within the one-year period, dynamically capturing claim inflation and tail development. Such precise and dynamic estimation of IBNR proportions is essential for maintaining premium adequacy and effectively managing future solvency risk.

### 6.2 Methodology

In this section, we consider the total cost of claims from the incurred but not reported exposure over the one-year projection horizon. The ATRP simulation involves:

- Claim Occurrence Modeling: Claims incurred within the one-year horizon are simulated using the ATRP calibrated with the trend function below

$$\lambda(t) = p_1 \frac{\lambda_1^{\alpha_1} t^{\alpha_1 - 1}}{\Gamma(\alpha_1)} e^{-\lambda_1 t} + p_2 \frac{\lambda_2^{\alpha_2} t^{\alpha_2 - 1}}{\Gamma(\alpha_2)} e^{-\lambda_2 t},$$

  where $t$ is in days.

- Claim Cost Assignment: Each simulated claim is paired with a severity draw from the fitted payment distribution, adjusted for inflation.

- Exposure Estimation: Instead of assuming a fixed IBNR share (e.g., 50%), we estimate the effective IBNR proportion using $PIBNR_{N(t)}$ and $PIBNR_{Z(t)}$ as defined below.

It is important to distinguish between the structural framework and the numerical projection engine. The micro-level reserve structure does not intrinsically depend on a trend renewal process. In the present study, the ATRP is used to generate internally consistent future claim trajectories so that RBNS, IBNR, and UPR can all be evaluated under a common probabilistic mechanism. The IBNR proportions obtained in this section are therefore supported by empirical calibration of reporting and settlement delays on the Florida dataset, and not solely driven by hypothetical simulation paths.

We define the Claim-Count-Based IBNR proportion as

$$PIBNR_{N(t)} = \frac{E[N_{tc}(t)]}{E[N_{occ}(t)]},$$

where

$$N_{occ}(t) = \sum_{k=1}^{N(t)} 1_{\{T_k \leq t\}}$$

and

$$N_{tc}(t) = \sum_{k=1}^{N(t)} 1_{\{T_k + \xi_k > t\}}.$$

$N_{tc}(t)$ is the number of claims of the tail coverage. $N_{occ}(t)$ is the number of claims of the occurrence coverage. We also define the Claim-Cost-Based IBNR (proportion) as

$$PIBNR_{Z(t)} = \frac{E[Z_{tc}(t)]}{E[Z_{occ}(t)]},$$

where

$$Z_{occ}(t) = \sum_{k=1}^{N(t)} 1_{\{T_k \leq t\}} \{A_1(T_k + \xi_k + \zeta_k) X_k + A_2(T_k + \xi_k + \zeta_k) Y_k\}$$

and

$$Z_{tc}(t) = \sum_{k=1}^{N(t)} 1_{\{T_k + \xi_k > t\}} \{A_1(T_k + \xi_k + \zeta_k) X_k + A_2(T_k + \xi_k + \zeta_k) Y_k\}.$$

$Z_{occ}(t)$ is the cost of the occurrence coverage. $Z_{tc}(t)$ is the cost of the tail coverage.

The required quantities—representing counts and costs of tail and occurrence claims—are estimated within the proposed unified micro-level framework using a specific stochastic implementation for event timing and cost dynamics. While the current implementation relies on an ATRP-based simulation for coherence with the reserving structure, alternative modeling assumptions and process specifications could also be considered. The framework nevertheless yields realistic estimates by appropriately capturing claim timing and inflation adjustments.

### 6.3 IBNR Proportions Estimates under Varying Trend Assumptions with an Arbitrary Trend (One-year horizon)

This section presents the IBNR proportions estimates computed under three distinct values of the trend parameter $\gamma$, which governs the intensity and severity of claim emergence over time. Specifically, we consider values of $\gamma = 0.5, 1, 1.5$, corresponding respectively to mild, moderate, and strong underlying trends in the occurrence process. The simulation framework produces IBNR proportions for both claim-count-based and claim-cost-based estimators, as summarized in Table 22.

Table 22
IBNR Proportions under Baseline Inflation (0%)

| Trend | IBNR (Count-Based) | IBNR (Cost-Based) |
|---|---|---|
| $\gamma = 0.5$ | 0.658 | 0.651 |
| $\gamma = 1.0$ | 0.733 | 0.727 |
| $\gamma = 1.5$ | 0.780 | 0.775 |

The proportions indicate a clear positive relationship with the trend parameter. Higher trends correspond to greater IBNR liabilities, reflecting the ATRP model's ability to capture delayed emergence and escalating severity. Both count-based and cost-based methods yield similar patterns, with cost-based estimates slightly elevated due to inflation effects.

These findings highlight the flexibility and robustness of our framework in estimating IBNR proportions and underscore the importance of dynamic rather than static trend assumptions.

### 6.4 Sensitivity to Inflation Shocks with an Arbitrary Trend (One-Year Horizon)

Although the qualitative direction of several sensitivities—such as the increase of IBNR with longer reporting or settlement delays—is theoretically predictable, their quantitative magnitude is not. From a practical actuarial standpoint, capital requirements and pricing margins respond to the magnitude of these effects rather than to their sign alone. The sensitivity analyses in this section are therefore designed not to establish directional effects, but to quantify their economic materiality under realistic calibration.

Therefore, to evaluate the robustness of IBNR proportion estimates under varying inflation assumptions, we consider a range of deterministic inflation shocks applied to the severity component of the model. Specifically, we examine four scenarios relative to the baseline (0% inflation): a +1% shock, a −1% shock, and more extreme ±5% shocks. The resulting proportions are reported in Tables 23 through 26 for both claim-count-based and cost-based measures.

The IBNR proportions under each case are reported below for both count-based and cost-based approaches.

Table 23
IBNR Proportions under +1% Inflation Shock

| Trend | IBNR (Count-Based) | IBNR (Cost-Based) |
|---|---|---|
| $\gamma = 0.5$ | 0.658 | 0.654 |
| $\gamma = 1.0$ | 0.733 | 0.731 |
| $\gamma = 1.5$ | 0.780 | 0.778 |

Table 24
IBNR Proportions under −1% Inflation Shock

| Trend | IBNR (Count-Based) | IBNR (Cost-Based) |
|---|---|---|
| $\gamma = 0.5$ | 0.658 | 0.647 |
| $\gamma = 1.0$ | 0.733 | 0.724 |
| $\gamma = 1.5$ | 0.780 | 0.772 |

Table 25
IBNR Proportions under +5% Inflation Shock

| Trend | IBNR (Count-Based) | IBNR (Cost-Based) |
|---|---|---|
| $\gamma = 0.5$ | 0.658 | 0.670 |
| $\gamma = 1.0$ | 0.734 | 0.743 |
| $\gamma = 1.5$ | 0.780 | 0.789 |

Table 26
IBNR Proportions under −5% Inflation Shock

| Trend | IBNR (Count-Based) | IBNR (Cost-Based) |
|---|---|---|
| $\gamma = 0.5$ | 0.658 | 0.633 |
| $\gamma = 1.0$ | 0.734 | 0.712 |
| $\gamma = 1.5$ | 0.780 | 0.762 |

The count-based IBNR proportions remain invariant across inflation scenarios, consistent with the model structure wherein claim frequency is independent of cost dynamics. By contrast, cost-based IBNR proportion estimates are sensitive to inflation, with upward shocks yielding proportionally higher reserves. This divergence becomes more pronounced as the trend parameter $\gamma$ increases, reflecting a compounding interaction between inflation and the tail distribution of claim severity.

Under moderate conditions (±1% inflation), the effect on cost-based IBNR is modest, indicating stability in reserve proportions for small cost perturbations. However, at ±5%, deviations in IBNR (cost-based) estimates become non-negligible—on the order of 3 to 5 percentage points—suggesting that high-inflation environments can materially affect reserve adequacy, particularly for portfolios with strong upward cost trends.

These results underscore the importance of explicitly incorporating inflation assumptions in liability valuation models. While minor inflation fluctuations may be absorbed by existing margins, sustained or extreme inflation shocks can lead to an underestimation of reserve needs if not properly reflected in the underlying actuarial assumptions.

**6.5 IBNR Estimates Using Calibrated Trend from 2014 Data (One-year horizon)**

In this section, we further illustrate the flexibility of our framework by calibrating the trend on data available up to 2015 and then applying it to forecast IBNR for subsequent policy periods (2014–2015 and 2015–2016).

We simulate future claims over the unearned portion of coverage for the two projection horizons and compute the IBNR ratios for both count- and cost-based measures. The results are presented in Table 27.

We estimate the trend using the approach from Léveillé and Hamel (2017). The graph of the calibrated trend is shown in Figure 2. The trend has the following functional form:

$$\lambda(t) = p_1 \frac{\lambda_1^{\alpha_1} t^{\alpha_1-1}}{\Gamma(\alpha_1)} e^{-\lambda_1 t} + p_2 \frac{\lambda_2^{\alpha_2} t^{\alpha_2-1}}{\Gamma(\alpha_2)} e^{-\lambda_2 t}$$

where $t$ is expressed in days.

Table 27
IBNR Proportions for Calibrated Trend

| Period | IBNR (Count-Based) | IBNR (Cost-Based) |
|---|---|---|
| 2014-2015 | 0.71737057 | 0.71267819 |
| 2015-2016 | 0.71607187 | 0.71050231 |

These results are obtained using the calibrated function $\lambda(t)$, described above. These findings confirm the stability of IBNR proportions under the calibrated trend, with both count-based and cost-based estimates consistently near 71–72%. This demonstrates that even without access to updated premiums for 2015 and beyond, the ATRP model can produce reasonable forward-looking IBNR estimates based on historically inferred trend and occurrence dynamics.

Practically, this supports the use of estimated trend parameters in reserving exercises, particularly in contexts where portfolio data is incomplete or corporate transactions interrupt the availability of premium information.

## 6.6 Sensitivity of IBNR Estimates to Reporting and Settlement Delays with Calibrated Trend

An important feature of our framework for IBNR estimation is its ability to disentangle and quantify the separate effects of claim reporting and settlement processes. To demonstrate this, we conduct a sensitivity analysis whereby the assumed average reporting and settlement delays are systematically halved and doubled, while holding all other model parameters fixed. For each scenario, we compute the IBNR proportions based on both claim counts and claim costs, for two successive policy years (2014–2015 and 2015–2016). The results are summarized in Table 28.

Table 28
IBNR Proportions Under Different Delay Scenarios

| Scenario | Year | IBNR (Count-based) | IBNR (Cost-based) |
|---|---|---|---|
| No Change | 2014-2015 | 0.71737057 | 0.71267819 |
| | 2015-2016 | 0.71607187 | 0.71050231 |
| Double Reporting | 2014-2015 | 0.84143893 | 0.83635967 |
| | 2015-2016 | 0.84052208 | 0.83479666 |
| Half Reporting | 2014-2015 | 0.52005031 | 0.5163746 |
| | 2015-2016 | 0.51777628 | 0.51404511 |
| Double Settlement | 2014-2015 | 0.71737057 | 0.71267114 |
| | 2015-2016 | 0.71607187 | 0.71067908 |
| Half Settlement | 2014-2015 | 0.71737057 | 0.71269424 |
| | 2015-2016 | 0.71607187 | 0.71039439 |

Across all scenarios, the IBNR estimates exhibit a marked sensitivity to changes in reporting delay, whereas the impact of settlement delay variations is negligible. Specifically, doubling the reporting delay results in an increase of the IBNR proportion by approximately 17% relative to the baseline scenario. Conversely, halving the reporting delay reduces the IBNR proportion by over 25%. In contrast, halving or doubling settlement delays has virtually no discernible effect on the IBNR estimates, which remain stable to within two decimal places.

This asymmetry between the two types of operational delays is both theoretically and practically justified. From a theoretical standpoint, IBNR proportion reflects the portion of incurred corresponding to the yet-reported at the valuation date. As such, it is directly linked to the timing of when insured events are reported, since only reported claims are reflected in the recognition of incurred losses and reserves. Delays in settlement, by contrast, affect the cash-flow profile of claims but do not alter the allocation of earned versus unearned premium, which depends solely on the exposure period and the recognition of reported losses. This distinction is well known in actuarial literature, yet to our knowledge, it has not been explicitly demonstrated in a stochastic claims process model with empirically calibrated parameters.

From an actuarial practice perspective, these findings highlight the critical role of accurately modeling reporting lags when estimating unearned premium reserves and assessing premium risk. In particular, practitioners should be aware that portfolios with longer expected reporting delays will exhibit higher IBNR proportions at any point during the policy period, implying a slower recognition of earned premium and a correspondingly higher future liability. Conversely, for portfolios with shorter reporting lags (such as certain property lines), the earning pattern accelerates, and IBNR diminishes accordingly. The negligible sensitivity to settlement delays underscores that while settlement speed is crucial for liquidity and claims management, it does not materially influence the accounting and reserving of unearned premiums.

By explicitly capturing the differential impact of reporting and settlement delays on IBNR, the proposed framework provides actuaries with a more granular and informative tool for premium risk assessment. It facilitates scenario testing, enabling practitioners to quantify the implications of operational changes (e.g., improved claims reporting systems) or shifts in portfolio mix on the reserve dynamics. These insights are especially relevant for insurers managing lines of business with heterogeneous reporting patterns, where traditional proportional methods of IBNR proportion estimation may overlook the operational nuances inherent in the claims process.

Overall, the sensitivity analysis underscores the practical applicability of the ATRP approach in providing a theoretically sound and empirically validated mechanism to support both regulatory reporting and internal risk management functions. By linking IBNR behaviour explicitly to reporting dynamics, our model enhances the actuary's ability to align reserving strategies with operational realities and to anticipate the effects of changes in claims handling practices on premium risk.

**6.7 UPR Estimates with Calibrated Trend (Six-Month Horizon)**

In this subsection, we shift our analytical focus from previously examined IBNR proportions to explicitly modeling Unearned Premium Reserves (UPR). While IBNR addresses claims incurred but not reported, UPR represents liabilities associated with premiums received for coverage periods extending into the future, capturing potential claim events that have not yet occurred. For practical relevance and to reflect realistic short-term insurance practice scenarios, we employ a six-month projection horizon for this UPR analysis.

To estimate UPR proportions, we utilize the calibrated ATRP, which accounts dynamically for both claim frequency and severity components. The calibrated trend function, derived from historical data up to the end of 2014, is employed consistently. Specifically, we analyze the sensitivity of UPR proportions to variations in operational delays—namely, claim reporting and settlement delays—thus providing additional insights into the underlying mechanics of premium earning and reserve allocation.

We define the Claim-Count-Based UPR proportion explicitly as follows:

$$PUPR_{N(t)}(t,h) = \frac{E[N_{occ}(t+h)] - E[N_{occ}(t)]}{E[N_{occ}(t+h)]},$$

where

$$N_{occ}(t) = \sum_{k=1}^{N(t)} 1_{\{T_k \leq t\}},$$

and $N_{occ}(t)$ is the number of claims of the occurrence coverage. We define the Claim-Cost-Based UPR proportion as

$$PUPR_{Z(t)}(t,h) = \frac{E[Z_{occ}(t+h)] - E[Z_{occ}(t)]}{E[Z_{occ}(t+h)]},$$

where

$$Z_{occ}(t) = \sum_{k=1}^{N(t)} 1_{\{T_k \leq t\}} \{A_1(T_k + \xi_k + \zeta_k)X_k + A_2(T_k + \xi_k + \zeta_k)Y_k\}.$$

$Z_{occ}(t)$ is the cost of the occurrence at time $t = 0.5$ (half a year). Moreover, we let $h = 0.5$ and $E[Z_{occ}(t+h)]$ is the value at $t + h = 1$ (one year).

Table 29 below summarizes UPR proportions under varying operational delay scenarios, including the baseline ("No Change"), double and half reporting delays, and double and half settlement delays. The calibrated trend function remains consistent throughout all scenarios.

Table 29
UPR Proportions Under Different Delay Scenarios (Six-Month Horizon)

| Scenario | Year | UPR (Count-based) | UPR (Cost-based) |
|---|---|---|---|
| No Change | 2014-2015 | 0.448277893 | 0.44725559 |
| | 2015-2016 | 0.44252409 | 0.44026581 |
| Double Reporting | 2014-2015 | 0.44827789 | 0.44731789 |
| | 2015-2016 | 0.44252409 | 0.44027181 |
| Half Reporting | 2014-2015 | 0.44827789 | 0.44722401 |
| | 2015-2016 | 0.44252409 | 0.44026243 |
| Double Settlement | 2014-2015 | 0.44827789 | 0.44711058 |
| | 2015-2016 | 0.44252409 | 0.4400521 |
| Half Settlement | 2014-2015 | 0.44827789 | 0.44734136 |
| | 2015-2016 | 0.44252409 | 0.44046572 |

A careful examination of Table 29 reveals several crucial insights. First, both count-based and cost-based UPR proportions exhibit remarkable stability across all delay scenarios over the six-month horizon. This observation indicates that, in contrast to the marked sensitivity previously observed for IBNR proportions with respect to reporting delays, the shorter horizon and nature of UPR (related to future claims yet to occur) inherently insulate these measures from significant impacts due to operational delay modifications. The count-based UPR remains effectively constant across all scenarios, consistently near approximately 44.8% and 44.3% for policy years 2014–2015 and 2015–2016, respectively.

The cost-based UPR shows slightly more variability but remains tightly clustered with the count-based results, typically differing only at the third decimal point. This minor variation is attributable to the cost-based measure's direct sensitivity to changes in claim inflation assumptions implicitly affected by settlement timing. Interestingly, doubling or halving the settlement delay has a modest, though discernible, influence on cost-based UPR proportions—reflecting subtle differences in the discounted expected value of future claim costs due to shifts in cash-flow timing. However, the magnitudes of these changes remain practically negligible (generally within 0.05 percentage points), reinforcing the overall robustness of UPR estimates within a short-term horizon.

The negligible effect of varying reporting delays on UPR further highlights the distinct conceptual differences between UPR and IBNR measures. Because UPR captures liability exposure for events yet to occur, the timing of claim reporting has no direct bearing on its estimation. Conversely, settlement delays impact only the discounted cost estimates rather than the exposure itself. This theoretical distinction is explicitly borne out by our empirical analysis.

From a practical actuarial perspective, these results underline the reliability of short-term UPR estimates provided by the proposed framework. The demonstrated stability and robustness to delay scenarios are particularly reassuring for actuaries seeking accurate, actionable reserve estimates for solvency management and premium adequacy analyses over shorter forecasting periods. In a dynamic insurance market, where operational practices such as claims processing efficiency and payment timing can vary, our framework's ability to produce stable reserve proportions enhances its utility and credibility as a tool for internal risk assessment and regulatory reporting.

In summary, our comprehensive analysis in this subsection demonstrates that the proposed framework yields highly stable, practically robust short-term UPR estimates. These findings complement and enrich our earlier analyses of IBNR proportions, emphasizing the versatility and effectiveness of ATRP modeling across different reserve categories, operational environments, and time horizons.

## 6.8 Sensitivity of UPR Estimates to Inflation Shocks under Varying Trend Assumptions with an Arbitrary Trend (Six-month Horizon)

This subsection rigorously investigates how Unearned Premium Reserve (UPR) estimates respond to simultaneous variations in inflation assumptions and the underlying claim occurrence trend within a six-month projection horizon. Recognizing the crucial impact of claim trends and inflation dynamics on reserving accuracy—particularly within long-tailed business lines—we systematically analyze UPR proportions under multiple trend scenarios and inflation shocks. Specifically, we evaluate three distinct scenarios of the trend parameter $\gamma$, each representing different levels of expected claim frequency and severity escalation:

- Mild trend ($\gamma = 0.5$): Reflecting a relatively modest increase in claim occurrence frequency and severity, typically representing stable or moderately evolving risk environments.

- Moderate trend ($\gamma = 1$): Indicative of more pronounced changes in claim patterns, corresponding to intermediate levels of frequency or severity escalation.

- Strong trend ($\gamma = 1.5$): Representing significant upward shifts in claim emergence and/or cost inflation, simulating adverse or rapidly deteriorating risk scenarios.

Table 30 presents the baseline UPR proportions under three arbitrarily selected values of the trend parameter: $\gamma = 0.5$ (mild trend), $\gamma = 1.0$ (moderate trend), and $\gamma = 1.5$ (strong trend), with no inflation applied. As expected, UPR proportions increase monotonically with the trend intensity, reflecting a greater share of future claims as the assumed growth in claim frequency and severity strengthens. Under baseline conditions, the count-based UPR rises from 0.290 to 0.647 as $\gamma$ increases from 0.5 to 1.5, while the corresponding cost-based UPR grows from 0.120 to 0.600. This consistent pattern is a direct consequence of the underlying ATRP dynamics, where increasing trend values elevate both the number and magnitude of future claims arising from unexpired coverage. The amplification of the cost-based UPR relative to the count-based version further underscores the compounding impact of severity inflation embedded in the claim generation process.

Table 30
UPR under Baseline Inflation (0%) and Varying Trend Assumptions

| Trend | UPR (Count-Based) | UPR (Cost-Based) |
|---|---|---|
| $\gamma = 0.5$ | 0.290 | 0.120 |
| $\gamma = 1.0$ | 0.501 | 0.415 |
| $\gamma = 1.5$ | 0.647 | 0.600 |

To assess the robustness of these estimates, we apply deterministic shocks of $\pm 1\%$ and $\pm 5\%$ to the inflation rate. Tables 31 through 34 report the UPR proportions under these perturbed inflation scenarios for each trend level. The count-based UPR remains unchanged across all inflation shocks, highlighting the independence between claim frequency and inflation assumptions in our model specification. By contrast, the cost-based UPR shows measurable sensitivity to inflation perturbations, with the response magnitude depending on the underlying trend. Under a +1% inflation shock, the cost-based UPR increases slightly across all trend levels (e.g., from 0.600 to 0.602 at γ = 1.5), while a −1% shock yields symmetric downward shifts. Larger shocks of ±5% yield more pronounced deviations. For instance, under the mild trend scenario, the cost-based UPR increases from 0.120 to 0.139 with a +5% inflation shock and decreases to 0.101 with a −5% shock. At higher trend intensities, the absolute impact of inflation shocks increases, but their relative effect diminishes as the baseline cost-based UPR grows.

Table 31
UPR under +1% Inflation Shock

| Trend | UPR (Count-Based) | UPR (Cost-Based) |
|---|---|---|
| $\gamma = 0.5$ | 0.290 | 0.124 |
| $\gamma = 1.0$ | 0.501 | 0.418 |
| $\gamma = 1.5$ | 0.647 | 0.602 |

Table 32
UPR under -1% Inflation Shock

| Trend | UPR (Count-Based) | UPR (Cost-Based) |
|---|---|---|
| $\gamma = 0.5$ | 0.290 | 0.116 |
| $\gamma = 1.0$ | 0.501 | 0.413 |
| $\gamma = 1.5$ | 0.647 | 0.598 |

Table 33
UPR under +5% Inflation Shock

| Trend | UPR (Count-Based) | UPR (Cost-Based) |
|---|---|---|
| $\gamma = 0.5$ | 0.290 | 0.139 |
| $\gamma = 1.0$ | 0.501 | 0.429 |
| $\gamma = 1.5$ | 0.647 | 0.609 |

Table 34
UPR under −5% Inflation Shock

| Trend | UPR (Count-Based) | UPR (Cost-Based) |
|---|---|---|
| $\gamma = 0.5$ | 0.290 | 0.101 |
| $\gamma = 1.0$ | 0.501 | 0.402 |
| $\gamma = 1.5$ | 0.647 | 0.591 |

These results illustrate two important features of the ATRP-based reserve model. First, the trend parameter γ is a primary driver of unearned reserve proportions, governing the temporal distribution of future claims and thereby determining the fraction of risk that remains unearned at any valuation point. The ability of the model to replicate increasing UPR levels with escalating trend values confirms its responsiveness to structural changes in claim occurrence behavior.

Second, while inflation shocks do not influence claim counts, they significantly affect the present value of future claim costs, and this sensitivity is most apparent in scenarios with lower baseline reserves, where relative changes in cost projections are proportionally larger.

From an actuarial and regulatory standpoint, these findings reinforce the need for dynamic reserving approaches that jointly consider structural trend evolution and economic uncertainty. Static allocation heuristics—such as assuming a fixed percentage of written premium as unearned—are likely to misrepresent liability exposure in environments where claim trends or inflation are non-stationary. The suggested framework, by capturing both the structural drivers of claim development and the economic adjustments through inflation, provides a forward-looking and risk-sensitive tool for estimating unearned premium liabilities. The empirical patterns documented here suggest that even modest changes in inflation expectations can materially impact reserve estimates when compounded with strong underlying trends, thereby affecting solvency margins and risk capital assessments.

In sum, the sensitivity analysis in this subsection demonstrates the importance of explicitly incorporating trend and inflation interactions into unearned premium reserve calculations. The proposed framework offers a transparent and empirically grounded method for quantifying these interactions, enhancing both the interpretability and accuracy of premium liability estimates under diverse future scenarios.

## 6.9 Implications and Discussion

The comprehensive analyses presented throughout Section 6 underscore several substantive contributions to actuarial reserving practice, demonstrating the flexibility, robustness, and empirical utility of our proposed framework. Notably, this section has addressed two fundamental categories of future insurance liabilities: Incurred But Not Reported (IBNR) reserves, which pertain to claims already incurred but not yet observed, and Unearned Premium Reserves (UPR), which reflect exposure to claims yet to occur within the unexpired portion of active policies. By jointly modeling these distinct but complementary sources of liability within a unified stochastic framework, we operationalize the ATRP in a manner that aligns with both regulatory expectations and real-world reserving needs.

A central contribution of this section lies in its explicit dual application of the suggested framework to IBNR (Sections 6.1–6.6) and UPR (Sections 6.7–6.8), using both calibrated and arbitrary trend specifications. In the IBNR setting, a one-year projection horizon captures the emergence of delayed claims from already-covered exposure, and the model is calibrated on historical data up to year-end 2014. This calibration enables forward-looking estimation of IBNR proportions under conditions where future premiums and claims are unavailable, such as after a corporate portfolio transfer. In contrast, the UPR analysis focuses on a six-month horizon and addresses claims that have not yet occurred but will emerge from the unexpired coverage. Both analyses incorporate the same core model assumptions but apply them to different segments of the liability structure, reflecting the ATRP's adaptability to various reserving objectives.

Importantly, our results demonstrate that the ATRP is not merely referenced, but fully operationalized and empirically validated across multiple use cases. The claim occurrence process is simulated dynamically using trend functions derived either from calibration or plausible assumptions, and the corresponding claim costs are adjusted for inflation. This structure supports a broad range of forecasting and stress-testing exercises, enabling insurers to anticipate reserve behavior under changing operational and economic conditions.

The comparison between IBNR and UPR also reveals instructive differences in sensitivity to operational delays and inflation. IBNR proportions are highly sensitive to reporting delays, consistent with their dependence on the timing of incurred loss recognition. By contrast, UPR proportions—representing future claims—are effectively invariant to reporting delays and exhibit only minor sensitivity to settlement delays, especially over shorter horizons. Inflation shocks affect cost-based estimates in both settings, but the impact is more pronounced on the IBNR. These distinctions underscore the importance of matching modeling assumptions to the liability type and time horizon when designing reserve estimation frameworks.

Our use of both arbitrary trends (to test the model's structural sensitivity) and calibrated trends (to replicate historical reserving conditions) further illustrates the ATRP's practical versatility. Arbitrary trends facilitate scenario analysis and provide insight into how reserve adequacy responds to different claim emergence assumptions. Calibrated trends, on the other hand, enable empirical projections grounded in observed portfolio dynamics. The consistency of results across these implementations reinforces the model's internal coherence and its capacity to support diverse actuarial functions—including solvency assessment, pricing support, and capital planning.

Moreover, the inclusion of both count-based and cost-based estimation methods offers a richer and more nuanced perspective on premium risk. Across both IBNR and UPR estimations, these dual measures allow practitioners to examine not only how many claims may arise from unexpired or unreported exposure, but also how severe they are likely to be in monetary terms. The close alignment of count-based and cost-based UPR results, for example, highlights the model's stability in the short term, while the divergence observed in IBNR estimates under inflation shocks reveals important areas of risk concentration.

From a broader perspective, the proposed framework enhances the connection between reserving and pricing by explicitly linking the projection of future claims to dynamic exposure, operational timing, and economic conditions. This approach enables more granular risk segmentation, better alignment between premiums and expected liabilities, and improved communication with regulators and stakeholders regarding capital adequacy and solvency planning. As the insurance industry continues to evolve toward more forward-looking and risk-sensitive actuarial standards, models like the ATRP—which can flexibly accommodate complex operational realities and economic dynamics—will play a crucial role.

In summary, Section 6 demonstrates that our framework offers a unified, empirically grounded, and operationally viable tool for modeling both IBNR and UPR components of premium risk. It incorporates dynamic occurrence processes, inflation effects, reporting and settlement lags, and accommodates both historical calibration and scenario-based stress testing. By explicitly quantifying the sensitivity of future liabilities to key actuarial and economic drivers, the ATRP enables more robust reserve estimation, strengthens the integration of pricing and reserving functions, and supports the actuarial profession's evolving role in strategic risk management and regulatory compliance.

## 7. RISK CAPITAL IMPLICATIONS

The predictive distribution obtained from our conditional ATRP model (with parameters uncertainty) offers all possible ranges of reasonable reserves, but it is also essential from a risk capital standpoint. The predictive distribution enables the assessment of risk capital, where extreme loss events have to be considered for an insurance portfolio. Therefore, in addition to holding enough reserves, property and casualty (P&C) insurance companies have to hold risk capital relating to (unexpected) losses incurred but not yet paid to comply with regulatory standards. This amount is considered as a buffer to face adverse scenarios. In this section, we want to highlight the impact of inflation, discounting, dependence, and trend on the risk capital analysis.

The OSFI and AMF require Canadian insurance companies to set internal targets for risk capital. Mathematically, the risk capital is the difference between the risk measure and the expected unpaid losses of the portfolio. For the risk measure, we consider the tail value-at-risk (TVaR), also called expected shortfall (ES) or conditional tail expectation (CTE) (when applied to continuous random variables), which is believed to be more informative to actuaries than the value at risk (VaR) in the distribution tail, and the subadditivity of VaR is not guaranteed in general which means that VaR is then considered as a non-coherent risk measure. See Denuit et al. (2005) for further details on (coherent) risk measures.

The risk capital is defined as the difference between the risk measure and the value of liability (see, e.g., Dhaene et al. (2006)). The risk measure is used at a risk tolerance of 95% to replicate what is usually done in practice. At the same time, the value of liability (reserve) is generally assumed to be equal to the risk measure but at a risk tolerance set between 60% and 80%, according to the risk appetite. In fact, rather than relying on the expected (average) reserve value as in previous sections, we adopt the tail value-at-risk (TVaR) at specified confidence levels as the central risk measure. This choice is consistent with standard practice in risk management, where TVaR is preferred over the mean because it captures tail risk, is coherent (sub-additive), and reflects extreme but plausible outcomes that insurers need to withstand. Specifically, we use TVaR at the 60% level as a benchmark for reserves—reflecting a moderately prudent risk tolerance aligned with internal reserve-setting practices—and TVaR at the 95% level to represent a higher quantile used in capital adequacy testing under solvency regimes. This dual-level approach reflects both the operational perspective of reserving and the more conservative perspective of capital adequacy, ensuring that the analysis remains relevant to practitioners.

Mathematically, the risk capital associated with a risk R, noted by $\text{RC}(R)$, is then calculated as follows:

$$\text{RC}(R) = TVaR_{95\%}(R) - TVaR_{60\%}(R)$$

For risk capital analysis in this context, it is usual to consider the uncertainty of parameters. Through this approach, Table 35 presents a comparative study of the impact of average inflation on risk capital for different discount rates. It seems clear that the average effect of inflation on the risk capital is very high. When inflation is considered (while fixing the discount rate), the risk capital increases significantly, reflecting the inflation impact on both loss reserves and risk capital calculations.

We should note, however, that the inflation impact has to be mitigated in Table 35. We must be careful to overestimate the effect of inflation since it starts at time $T_0$ and is applied over a much more extended period than the discount. In the context of this study, it is, therefore, intuitively expected that inflation will have a more significant effect than discounting.

In addition, we consistently observe that the average inflation impact seems to be negatively correlated with the discount factor. Indeed, when the discount factor is taken from 0% to 6%, the relative average inflation impact decreases by about 2.7 %.

Also, independently from inflation, the average risk capital decreases when we increase the discount factor. When the discount factor is taken from 0% to 6% for the inflated losses, the risk capital decreases by 11.4% for the conditional ATRP model. For non-inflated losses, the average risk capital drops by 10.0% for the conditional ATRP model when the discount factor is taken from 0% to 6%. It is worth mentioning here that the impact of the discount factor is higher when inflation is considered.

TABLE 35
Average discount-inflation impact on risk capital

| ... | $\alpha_1 = 0.045692$, $\alpha_2 = 0.041744$ | | | $\alpha_1 = \alpha_2 = 0$ | | | ... |
|---|---|---|---|---|---|---|---|
| $\beta_1 = \beta_2$ | TVaR60 | TVaR95 | Risk Capital | TVaR60 | TVaR95 | Risk Capital | Inflation impact |
| 0% | 110,341,323 | 138,059,327 | 27,718,004 | 78,338,600 | 95,398,706 | 17,060,105 | 62.5% |
| 2% | 106,284,563 | 132,849,614 | 26,565,051 | 75,630,114 | 92,069,326 | 16,439,211 | 61.6% |
| 4% | 102,483,406 | 127,993,984 | 25,510,577 | 73,085,738 | 88,959,427 | 15,873,688 | 60.7% |
| 6% | 98,916,891 | 123,462,342 | 24,545,451 | 70,692,435 | 86,051,401 | 15,358,967 | 59.8% |

After reviewing the impact of the inflation and discounting on the risk capital, we examine next the effect of the dependence between the k-th claim paid, $X_k$, and the k-th expense paid, $Y_k$, as well as $\zeta_k$, which is the insurer's delay in paying the claim and expenses. The results are presented in Table 36 below.

TABLE 36

Impact of dependence between $X_k, Y_k$ and $\zeta_k$ on risk capital

| ... | $X_k, Y_k$ and $\zeta_k$ dependent | | | $X_k, Y_k$ and $\zeta_k$ independent | | | ... |
|---|---|---|---|---|---|---|---|
| $\beta_1 = \beta_2$ | TVaR60 | TVaR95 | Risk Capital | TVaR60 | TVaR95 | Risk Capital | Dep. impact |
| 0% | 110,341,323 | 138,059,327 | 27,718,004 | 87,760,132 | 107,528,604 | 19,768,471 | 40.2% |
| 2% | 106,284,563 | 132,849,614 | 26,565,051 | 84,821,492 | 103,842,816 | 19,021,324 | 39.7% |
| 4% | 102,483,406 | 127,993,984 | 25,510,577 | 82,058,377 | 100,396,180 | 18,337,803 | 39.1% |
| 6% | 98,916,891 | 123,462,342 | 24,545,451 | 79,456,754 | 97,167,641 | 17,710,887 | 38.6% |

We observe that risk capital is greater when dependence is incorporated. The dependence impact is about 40% when we consider a dependence relation between the indemnities, expenses, and payment delays. This is unsurprising since we expect a heavier tail when considering this positive dependence. This also confirms the importance of adequately capturing this dependence, which helps strategic decisions, notably from a quantitative risk management standpoint.

Moreover, Table 36 affirms the influence of discounting on risk capital. Specifically, the impact of dependence on risk capital diminishes with an increase in the discount rate. As observed previously, an increasing discount rate correlates with reducing risk capital, both with and without dependence. Notably, when dependence is factored in, the risk capital for the conditional ATRP model decreases by 11.4% as the discount factor transitions from 0% to 6%. Relaxing the dependence assumption results in an average risk capital decrease of 10.4% for the conditional ATRP model under the same discount factor range. Interestingly, the discount factor's impact is marginally more significant when considering dependence.

In addition to inflation, discounting, and dependence, one can examine the impact on the risk capital (with parameters uncertainty) if we change the distribution of $\zeta_k$. Table 37 shows the (average) results obtained from 100,000 scenarios, as defined in Section 4.3.1.

TABLE 37

Impact on risk capital of a change in settlement delay $\zeta_k$

| ... | Calibrated $\zeta_k$ | | | New delay $\zeta_k$ | | | ... |
|---|---|---|---|---|---|---|---|
| $\beta_1 = \beta_2$ | TVaR60 | TVaR95 | Risk Capital | TVaR60 | TVaR95 | Risk Capital | S. D. Impact |
| 0% | 110,341,323 | 138,059,327 | 27,718,004 | 98,295,572 | 122,277,492 | 23,981,920 | - 13.5% |
| 2% | 106,284,563 | 132,849,614 | 26,565,051 | 96,192,990 | 119,599,628 | 23,406,638 | - 11.9% |
| 4% | 102,483,406 | 127,993,984 | 25,510,577 | 94,178,914 | 117,048,297 | 22,869,383 | - 10.4% |
| 6% | 98,916,891 | 123,462,342 | 24,545,451 | 92,247,873 | 114,615,145 | 22,367,272 | - 8.9% |

Indeed, when the calibrated delay $\zeta_k$ is replaced with the newly defined delay, as described earlier, the (average) risk capital decreases by approximately 9% to 13% with an increasing discount rate. Notably, there is a positive correlation between the impact of settlement delay on risk capital and the discount factor. Specifically, as the discount rate increases from 0% to 6%, the risk capital decreases by 11.4% for the calibrated delay $\zeta_k$ and by 6.7% for the new delay.

## 8. ABOUT THE TREND RENEWAL PROCESS

In this section, we discuss the potential for applying trend renewal processes to settlement delay trends, reinforcing the broader applicability of our methodology. This demonstrates that the trend renewal process can also be used to model RBNS claims to capture settlement delays, capturing their evolving patterns over time. Even if it was not necessary to know the distribution functions of $T_k$ and $\xi_k$ in this study, the different delays could generate trend renewal processes. It is a reasonable assumption in many actuarial contexts, and it is often implicitly incorporated in many datasets. Lindqvist et al. (2003) show how to make a statistical inference on a dataset to implement a TRP. In Léveillé and Hamel (2017), a discussion on an estimation method is presented by the authors to get an approximation of the trend function of the sequence of occurrence times $\{T_k; k \in N^*\}$ of medical malpractice "accidents". The counterpart of the TRP is that the inter-occurrence times of the resulting process are not generally iid, which implies many more calculations.

Moreover, let us point out that our trend functions could be used not only to get a deeper insight into the historical data but also be useful for examining the sensitivity of the reserve to different historical data if we use our ATRP model (with the calibrated delays) as a generating process of our dataset. However, as this last point is not the main objective of this paper and the examination of many trend functions would represent a cumbersome exercise, we will limit ourselves here to studying how to get a trend renewal distribution for the settlement delay (other than a renewal one).

Hence, assume that $\{\zeta_k; k \in N^*\}$ is a TRP$[F, \lambda(.)]$, where $\lambda(t)$ is an estimated or an arbitrarily chosen trend function and $F(t)$ is the (parametric) distribution function of the inter-occurrence times of the corresponding renewal process $\{\Lambda(\psi_k) - \Lambda(\psi_{k-1}); k \in N^*\}$, where $\psi_k = \zeta_1 + ... + \zeta_k$ and $\Lambda(t)$ being the cumulative trend function defined in Section 2. Then, the distribution functions of $\psi_k$ and $\zeta_k$ will be given by

$$F_{\psi_k}(t) = F_{\tilde{\psi}_k}(\Lambda(t)) = F^{*k}(\Lambda(t)) \quad , \quad F_{\zeta_k}(t) = \int_0^\infty F(\Lambda(u+t) - \Lambda(u)) dF^{*(k-1)}(\Lambda(u)).$$

If we use simulation (or our formulas) to estimate (or evaluate) the first two moments of our conditional ATRP model, we need to assess (or evaluate) the conditional distribution function of $\zeta_k | \zeta_k > t - t_k - \xi_k$ for each $k$, bearing in mind that these $\zeta_k$ could be dependent (if the trend function is not constant). Therefore, introducing a TRP model through numerical integration or simulation will generally involve more than through simulation.

To illustrate the use of a TRP$[F, \lambda(.)]$ for $\{\zeta_k; k \in N^*\}$ in our predictions, we apply this model to the settlement delays of the claims not settled in the first ten years of the accident. We have chosen the trend function, $\lambda(t) = 1.2 \, t^{0.2}$ (which implies $\Lambda(t) = t^{1.2}$), and the calibrated generalized gamma of section 3.2 as the distribution $F$ of the (virtual) inter-occurrence times. As in Section 4.3, this can be interpreted as a speed-up of the settlement delays to reduce the reserve. All the other calibrated distributions of Section 3.2 remain the same. Hereafter, in Table 38, the results were obtained from 100,000 simulations (without parameters' uncertainty).

Table 38
$\{\zeta_k; k \in N^*\}$: TRP$\left[F, \lambda(t) = 1.2 \, t^{0.2}\right]$

| $\beta_1 = \beta_2$ | Mean | St. dev. | VaR60 | VaR80 | VaR95 | TVaR60 | TVaR80 | TVaR95 |
|---|---|---|---|---|---|---|---|---|
| 0% | 71,437,349 | 8,392,230 | 72,485,992 | 77,6083,26 | 86,135,065 | 79,354,982 | 83,875,451 | 92,632,030 |
| 2% | 71,127,416 | 8,354,321 | 72,170,717 | 77,265,186 | 85,753,230 | 79,009,638 | 83,509,729 | 92,227,174 |
| 4% | 70,820,424 | 8,317,021 | 71,864,090 | 76,931,662 | 85,391,333 | 78,667,774 | 83,147,806 | 91,826,814 |
| 6% | 70,516,319 | 8,280,308 | 71,556,169 | 76,602,171 | 85,023,101 | 78,329,324 | 82,789,591 | 91,430,685 |

As usual, all the selected quantities are decreasing as the discount rates increase. The CVs are nearly the same for these discount rates (11.7%), and then the means remain representative regardless of the discount rates. For our discount rates going from 0% to 6%, all the selected quantities (means, st. dev., …) are each reduced approximatively by 1.3%. If you compare Table 38 to Table 10, for our discount rates going from 0% to 6%, the means are respectively reduced by 22.2%, 19.6%, 17.1%, 14.5% and the standard deviations by 17.2%, 14.1%, 11.1%, 8.2%. The decrease is between 13% and 23% for the other quantities. Moreover, if we now compare Table 38 to Table 16, for our discount rates going from 0% to 6%, the means are respectively reduced by 14.9%, 13%, 11.1%, 9.3% and the standard deviations by 10.1%, 8.0%, 6.1%, 4.3%. For the other quantities, the decrease is between 8.1% and 15.5%. Thus, we see that this TRP has reduced the settlement delays more meaningfully than a simple change of parameter in the generalized gamma distribution as used in Section 4.3.

**Remark 6.** (1) Another choice of the trend function could have been made but the selection of the generalized gamma as our distribution function $F$ was a more "natural" choice since it was calibrated on our historical data and we wanted to "perturbate" in some way this distribution for the non settled claims.

(2) It would have also been possible to calibrate a TRP model on our historical (virtual) settlement delays and use it to simulate our predictions.

# 9 CONCLUDING REMARKS

Micro-level reserving models offer considerable advantages when detailed, granular claim-level data are available, a situation increasingly common across various insurance lines. In this paper, we introduced a conditional Aggregate Trend Renewal Process (ATRP) model that integrates a range of structural components, including a trend renewal process for modeling accident counts, inflation-adjusted claim severities, delays in reporting and settlement, dependency structures between expenses and indemnities, and economic discounting. This multi-faceted structure allows for a richer representation of the complex dynamics underlying real-world reserving.

A notable innovation of our work is the full operationalization of the suggested framework across distinct actuarial applications. Beyond RBNS estimation, we extended the model to forecast liabilities associated with both Incurred But Not Reported (IBNR) and Unearned Premium Reserves (UPR). These two categories—corresponding to claims that have occurred but are not yet reported, and to future claims associated with unexpired coverage—are jointly modeled using the same stochastic process but under different temporal and structural assumptions. The IBNR analysis focuses on a one-year horizon, while the UPR estimation applies a six-month projection, consistent with short-term premium risk assessments. This dual application highlights the flexibility of the proposed architecture, addressing both incurred and future exposure, a feature rarely achieved within a single modeling platform.

Importantly, we implemented both calibrated trends, derived from historical data using the methodology of Léveillé and Hamel (2017), and arbitrary trend scenarios to assess sensitivity and stress-test model outputs. These two approaches enable a balance between empirical accuracy and forward-looking risk assessment. By using a calibrated trend to simulate realistic projections under incomplete data (e.g., following a company sale) and an arbitrary trend to examine structural sensitivity, we showed that our modeling setup can be adapted to both data-rich and data-limited environments.

Further refinements were achieved by incorporating covariates, notably claim type (e.g., injury category), into the severity modeling. This heterogeneity adjustment significantly enhanced predictive accuracy and stability across model runs. We demonstrated the capacity of the proposed micro-level modeling structure to produce reliable estimates of the first two moments of cell-level and aggregated reserves, approximate reserve distributions, and quantify uncertainty through sensitivity and scenario testing. In particular, the impact of operational features such as reporting and settlement delays, trend intensities, and deterministic inflation shocks were explicitly modeled and quantified. Taken together, these features position the proposed unified reserving approach as a robust and interpretable tool for reserve estimation.

Through our analysis, we also emphasized the actuarial significance of UPR as a distinct source of premium risk. Often treated simplistically in reserving practice, UPR reflects the unexpired risk on active policies and is highly relevant for premium adequacy, solvency margins, and risk-based capital requirements. Our model demonstrated that UPR estimates are sensitive to trend assumptions and inflationary environments but notably stable with respect to reporting delays—distinguishing them from IBNR estimates, which are primarily driven by unobserved reporting lag. This contrast offers practical guidance for insurers on which operational levers are most impactful depending on the nature of the reserve being estimated.

Despite its strengths, one limitation of the dataset used is the assumption of a single payment per claim. While this constraint reflects the structure of the available data, the proposed micro-level modeling structure could readily be adapted to accommodate multiple payments per claim where such data exist. Future work could explore this generalization to further enhance realism. More broadly, while the current implementation relies on a specific stochastic specification, alternative modeling assumptions and process formulations could also be considered within the same unified reserving architecture. Similarly, although claim type was used as a key covariate, incorporating additional predictors such as geographic location, policyholder demographics, or line-specific attributes could yield further gains in predictive power and actuarial relevance. These extensions would be particularly valuable in datasets capturing richer policy- and claim-level features.

Looking ahead, the ATRP model's structure is readily extendable to a wide range of insurance contexts beyond medical malpractice, including automobile, homeowners', and professional liability insurance. Introducing stochastic inflation and discount rates could allow the model to reflect more realistically the financial environment, while relaxing independence assumptions between severity and delay components would support more complex multivariate formulations. Such enhancements would further solidify the ATRP's applicability to modern actuarial challenges, particularly in regimes of increasing uncertainty and regulatory scrutiny.

In conclusion, this paper proposes a unified structure for the joint modeling of RBNS, IBNR, and unearned premium risk under dependence, inflation, and discounting. While an Aggregate Trend Renewal Process (ATRP) is employed in this study as an implementation device, the methodological contribution lies in the integrated reserving architecture itself, rather than in any specific renewal specification. The results demonstrate that the proposed framework can simultaneously capture payment dynamics, premium risk, operational delays, and dependence effects within a single coherent probabilistic structure. To the best of our knowledge, no existing model in the actuarial literature jointly addresses all of these dimensions at the micro level within a unified framework. This unification enables consistent forward-looking assessment of liabilities and premium risk, with direct implications for reserving practice, capital allocation, and solvency analysis. By bridging modeling components that are traditionally treated separately, the proposed approach offers a flexible and extensible foundation for both methodological advancement and practical decision-making in long-tailed lines of business.

**Data availability statement.** The data and code supporting this study's findings are available from the corresponding author, [AA], upon reasonable request.

**Funding statement.** This work received no specific grant from any funding agency, commercial or not-for-profit sectors.

**Competing interests.** The authors declare none.

## Appendix A

### - Definition of "ordinary renewal process"

A sequence of occurrence times $\{S_k; k \in \mathbf{N}^*\}$ is said to generate an ordinary renewal process if the sequence of the successive inter-occurrence times $\{\tau_k = S_k - S_{k-1}; k \in \mathbf{N}^*\}$, with $S_0 = 0$, are independent and identically distributed.

## Appendix B

### - Proof of Theorem 2  For identity (1), first and foremost, we have

$$E\left[W_{i,j}^2(t)\right] = E\left[\sum_{k=1}^{n_{B_i}} I_{]i+j-2, i+j-1]}\left(t_k^{B_i} + \xi_k^{B_i} + \zeta_k^{B_i}\right)\left\{A_1\left(t_k^{B_i} + \xi_k^{B_i} + \zeta_k^{B_i}\right)X_k^{B_i} + A_2\left(t_k^{B_i} + \xi_k^{B_i} + \zeta_k^{B_i}\right)Y_k^{B_i}\right\}\right.$$

$$\left. \times \sum_{r=1}^{n_{B_i}} I_{]i+j-2, i+j-1]}\left(t_r^{B_i} + \xi_r^{B_i} + \zeta_r^{B_i}\right)\left\{A_1\left(t_r^{B_i} + \xi_r^{B_i} + \zeta_r^{B_i}\right)X_r^{B_i} + A_2\left(t_r^{B_i} + \xi_r^{B_i} + \zeta_r^{B_i}\right)Y_r^{B_i}\right\}\Big| B_i\right]$$

$$= E\left[\sum_{k=1}^{n_{B_i}}\sum_{r=1}^{n_{B_i}} I_{]i+j-2, i+j-1]}\left(t_k^{B_i} + \xi_k^{B_i} + \zeta_k^{B_i}\right) I_{]i+j-2, i+j-1]}\left(t_r^{B_i} + \xi_r^{B_i} + \zeta_r^{B_i}\right)\right.$$

$$\left. \times A_1\left(t_k^{B_i} + \xi_k^{B_i} + \zeta_k^{B_i}\right) A_1\left(t_r^{B_i} + \xi_r^{B_i} + \zeta_r^{B_i}\right) X_k^{B_i} X_r^{B_i} \Big| B_i\right]$$

$$+ E\left[\sum_{k=1}^{n_{B_i}}\sum_{r=1}^{n_{B_i}} I_{]i+j-2, i+j-1]}\left(t_k^{B_i} + \xi_k^{B_i} + \zeta_k^{B_i}\right) I_{]i+j-2, i+j-1]}\left(t_r^{B_i} + \xi_r^{B_i} + \zeta_r^{B_i}\right)\right.$$

$$\left. \times A_1\left(t_k^{B_i} + \xi_k^{B_i} + \zeta_k^{B_i}\right) A_2\left(t_r^{B_i} + \xi_r^{B_i} + \zeta_r^{B_i}\right) X_k^{B_i} Y_r^{B_i} \Big| B_i\right]$$

$$+ E\left[\sum_{k=1}^{n_{B_i}}\sum_{r=1}^{n_{B_i}} I_{]i+j-2, i+j-1]}\left(t_k^{B_i} + \xi_k^{B_i} + \zeta_k^{B_i}\right) I_{]i+j-2, i+j-1]}\left(t_r^{B_i} + \xi_r^{B_i} + \zeta_r^{B_i}\right)\right.$$

$$\left. \times A_2\left(t_k^{B_i} + \xi_k^{B_i} + \zeta_k^{B_i}\right) A_1\left(t_r^{B_i} + \xi_r^{B_i} + \zeta_r^{B_i}\right) Y_k^{B_i} X_r^{B_i} \Big| B_i\right]$$

$$+ E\left[\sum_{k=1}^{n_{B_i}}\sum_{r=1}^{n_{B_i}} I_{]i+j-2, i+j-1]}\left(t_k^{B_i} + \xi_k^{B_i} + \zeta_k^{B_i}\right) I_{]i+j-2, i+j-1]}\left(t_r^{B_i} + \xi_r^{B_i} + \zeta_r^{B_i}\right)\right.$$

$$\left. \times A_2\left(t_k^{B_i} + \xi_k^{B_i} + \zeta_k^{B_i}\right) A_2\left(t_r^{B_i} + \xi_r^{B_i} + \zeta_r^{B_i}\right) Y_k^{B_i} Y_r^{B_i} \Big| B_i\right].$$

The first term yields,

$$E\left[\sum_{k=1}^{n_{B_i}}\sum_{r=1}^{n_{B_i}}I_{]i+j-2,i+j-1]}\left(t_k^{B_i}+\xi_k^{B_i}+\zeta_k^{B_i}\right)I_{]i+j-2,i+j-1]}\left(t_r^{B_i}+\xi_r^{B_i}+\zeta_r^{B_i}\right)\right.$$
$$\left.\times A_1\left(t_k^{B_i}+\xi_k^{B_i}+\zeta_k^{B_i}\right)A_1\left(t_r^{B_i}+\xi_r^{B_i}+\zeta_r^{B_i}\right)X_k^{B_i}X_r^{B_i}\Big|B_i\right]$$

$$=E\left[\sum_{k=1}^{n_{B_i}}I_{]i+j-2,i+j-1]}\left(t_k^{B_i}+\xi_k^{B_i}+\zeta_k^{B_i}\right)A_1^2\left(t_k^{B_i}+\xi_k^{B_i}+\zeta_k^{B_i}\right)\left(X_k^{B_i}\right)^2\right.$$
$$+2\sum_{k=1}^{n_{B_i}-1}\sum_{r=k+1}^{n_{B_i}}I_{]i+j-2,i+j-1]}\left(t_k^{B_i}+\xi_k^{B_i}+\zeta_k^{B_i}\right)I_{]i+j-2,i+j-1]}\left(t_r^{B_i}+\xi_r^{B_i}+\zeta_r^{B_i}\right)$$
$$\left.\times A_1\left(t_k^{B_i}+\xi_k^{B_i}+\zeta_k^{B_i}\right)A_1\left(t_r^{B_i}+\xi_r^{B_i}+\zeta_r^{B_i}\right)X_k^{B_i}X_r^{B_i}\Big|B_i\right].$$

The second term yields,

$$E\left[\sum_{k=1}^{n_{B_i}}\sum_{r=1}^{n_{B_i}}I_{]i+j-2,i+j-1]}\left(t_k^{B_i}+\xi_k^{B_i}+\zeta_k^{B_i}\right)I_{]i+j-2,i+j-1]}\left(t_r^{B_i}+\xi_r^{B_i}+\zeta_r^{B_i}\right)\right.$$
$$\left.\times A_1\left(t_k^{B_i}+\xi_k^{B_i}+\zeta_k^{B_i}\right)A_2\left(t_r^{B_i}+\xi_r^{B_i}+\zeta_r^{B_i}\right)X_k^{B_i}Y_r^{B_i}\Big|B_i\right]$$

$$=E\left[\sum_{k=1}^{n_{B_i}}I_{]i+j-2,i+j-1]}\left(t_k^{B_i}+\xi_k^{B_i}+\zeta_k^{B_i}\right)A_1\left(t_k^{B_i}+\xi_k^{B_i}+\zeta_k^{B_i}\right)A_2\left(t_k^{B_i}+\xi_k^{B_i}+\zeta_k^{B_i}\right)X_k^{B_i}Y_k^{B_i}\right.$$
$$+\sum_{k=1}^{n_{B_i}-1}\sum_{r=k+1}^{n_{B_i}}I_{]i+j-2,i+j-1]}\left(t_k^{B_i}+\xi_k^{B_i}+\zeta_k^{B_i}\right)I_{]i+j-2,i+j-1]}\left(t_r^{B_i}+\xi_r^{B_i}+\zeta_r^{B_i}\right)$$
$$\times A_1\left(t_k^{B_i}+\xi_k^{B_i}+\zeta_k^{B_i}\right)A_2\left(t_r^{B_i}+\xi_r^{B_i}+\zeta_r^{B_i}\right)X_k^{B_i}Y_r^{B_i}$$
$$+\sum_{r=1}^{n_{B_i}-1}\sum_{k=r+1}^{n_{B_i}}I_{]i+j-2,i+j-1]}\left(t_k^{B_i}+\xi_k^{B_i}+\zeta_k^{B_i}\right)I_{]i+j-2,i+j-1]}\left(t_r^{B_i}+\xi_r^{B_i}+\zeta_r^{B_i}\right)$$
$$\left.\times A_1\left(t_k^{B_i}+\xi_k^{B_i}+\zeta_k^{B_i}\right)A_2\left(t_r^{B_i}+\xi_r^{B_i}+\zeta_r^{B_i}\right)X_k^{B_i}Y_r^{B_i}\Big|B_i\right].$$

The third term yields,

$$E\left[\sum_{k=1}^{n_{B_i}}\sum_{r=1}^{n_{B_i}}I_{]i+j-2,i+j-1]}\left(t_k^{B_i}+\xi_k^{B_i}+\zeta_k^{B_i}\right)I_{]i+j-2,i+j-1]}\left(t_r^{B_i}+\xi_r^{B_i}+\zeta_r^{B_i}\right)\right.$$
$$\left.\times A_2\left(t_k^{B_i}+\xi_k^{B_i}+\zeta_k^{B_i}\right)A_1\left(t_r^{B_i}+\xi_r^{B_i}+\zeta_r^{B_i}\right)Y_k^{B_i}X_r^{B_i}\bigg|B_i\right]$$

$$=E\left[\sum_{k=1}^{n_{B_i}}I_{]i+j-2,i+j-1]}\left(t_k^{B_i}+\xi_k^{B_i}+\zeta_k^{B_i}\right)A_2\left(t_k^{B_i}+\xi_k^{B_i}+\zeta_k^{B_i}\right)A_1\left(t_k^{B_i}+\xi_k^{B_i}+\zeta_k^{B_i}\right)Y_k^{B_i}X_k^{B_i}\right.$$

$$+\sum_{k=1}^{n_{B_i}-1}\sum_{r=k+1}^{n_{B_i}}I_{]i+j-2,i+j-1]}\left(t_k^{B_i}+\xi_k^{B_i}+\zeta_k^{B_i}\right)I_{]i+j-2,i+j-1]}\left(t_r^{B_i}+\xi_r^{B_i}+\zeta_r^{B_i}\right)$$
$$\times A_2\left(t_k^{B_i}+\xi_k^{B_i}+\zeta_k^{B_i}\right)A_1\left(t_r^{B_i}+\xi_r^{B_i}+\zeta_r^{B_i}\right)Y_k^{B_i}X_r^{B_i}$$

$$+\sum_{r=1}^{n_{B_i}-1}\sum_{k=r+1}^{n_{B_i}}I_{]i+j-2,i+j-1]}\left(t_k^{B_i}+\xi_k^{B_i}+\zeta_k^{B_i}\right)I_{]i+j-2,i+j-1]}\left(t_r^{B_i}+\xi_r^{B_i}+\zeta_r^{B_i}\right)$$
$$\left.\times A_2\left(t_k^{B_i}+\xi_k^{B_i}+\zeta_k^{B_i}\right)A_1\left(t_r^{B_i}+\xi_r^{B_i}+\zeta_r^{B_i}\right)Y_k^{B_i}X_r^{B_i}\bigg|B_i\right].$$

The fourth term yields

$$E\left[\sum_{k=1}^{n_{B_i}}\sum_{r=1}^{n_{B_i}}I_{]i+j-2,i+j-1]}\left(t_k^{B_i}+\xi_k^{B_i}+\zeta_k^{B_i}\right)I_{]i+j-2,i+j-1]}\left(t_r^{B_i}+\xi_r^{B_i}+\zeta_r^{B_i}\right)\right.$$
$$\left.\times A_2\left(t_k^{B_i}+\xi_k^{B_i}+\zeta_k^{B_i}\right)A_2\left(t_r^{B_i}+\xi_r^{B_i}+\zeta_r^{B_i}\right)Y_k^{B_i}Y_r^{B_i}\bigg|B_i\right]$$

$$=E\left[\sum_{k=1}^{n_{B_i}}I_{]i+j-2,i+j-1]}\left(t_k^{B_i}+\xi_k^{B_i}+\zeta_k^{B_i}\right)A_2^2\left(t_k^{B_i}+\xi_k^{B_i}+\zeta_k^{B_i}\right)\left(Y_k^{B_i}\right)^2\right.$$

$$+2\sum_{k=1}^{n_{B_i}-1}\sum_{r=k+1}^{n_{B_i}}I_{]i+j-2,i+j-1]}\left(t_k^{B_i}+\xi_k^{B_i}+\zeta_k^{B_i}\right)I_{]i+j-2,i+j-1]}\left(t_r^{B_i}+\xi_r^{B_i}+\zeta_r^{B_i}\right)$$
$$\left.\times A_2\left(t_k^{B_i}+\xi_k^{B_i}+\zeta_k^{B_i}\right)A_2\left(t_r^{B_i}+\xi_r^{B_i}+\zeta_r^{B_i}\right)Y_k^{B_i}Y_r^{B_i}\bigg|B_i\right].$$

Adding the preceding terms, we get

$$E\left[W_{i,j}^2(t)\right] = E\left[\sum_{k=1}^{n_{B_i}} I_{]i+j-2,i+j-1]}\left(t_k^{B_i} + \xi_k^{B_i} + \zeta_k^{B_i}\right)\right.$$
$$\left. \times \left\{A_1^2\left(t_k^{B_i} + \xi_k^{B_i} + \zeta_k^{B_i}\right)\left(X_k^{B_i}\right)^2 + A_2^2\left(t_k^{B_i} + \xi_k^{B_i} + \zeta_k^{B_i}\right)\left(Y_k^{B_i}\right)^2\right\}\bigg|B_i\right]$$

$$+2E\left[\sum_{k=1}^{n_{B_i}-1}\sum_{r=k+1}^{n_{B_i}} I_{]i+j-2,i+j-1]}\left(t_k^{B_i} + \xi_k^{B_i} + \zeta_k^{B_i}\right)I_{]i+j-2,i+j-1]}\left(t_r^{B_i} + \xi_r^{B_i} + \zeta_r^{B_i}\right)\right.$$
$$\left. \times \left\{A_1\left(t_k^{B_i} + \xi_k^{B_i} + \zeta_k^{B_i}\right)A_1\left(t_r^{B_i} + \xi_r^{B_i} + \zeta_r^{B_i}\right)X_k^{B_i}X_r^{B_i} + A_2\left(t_k^{B_i} + \xi_k^{B_i} + \zeta_k^{B_i}\right)A_2\left(t_r^{B_i} + \xi_r^{B_i} + \zeta_r^{B_i}\right)Y_k^{B_i}Y_r^{B_i}\right\}\bigg|B_i\right\}\right]$$

$$+2E\left[\sum_{k=1}^{n_{B_i}} I_{]i+j-2,i+j-1]}\left(t_k^{B_i} + \xi_k^{B_i} + \zeta_k^{B_i}\right)A_1\left(t_k^{B_i} + \xi_k^{B_i} + \zeta_k^{B_i}\right)A_2\left(t_k^{B_i} + \xi_k^{B_i} + \zeta_k^{B_i}\right)X_k^{B_i}Y_k^{B_i}\bigg|B_i\right]$$

$$+2E\left[\sum_{k=1}^{n_{B_i}-1}\sum_{r=k+1}^{n_{B_i}} I_{]i+j-2,i+j-1]}\left(t_k^{B_i} + \xi_k^{B_i} + \zeta_k^{B_i}\right)I_{]i+j-2,i+j-1]}\left(t_r^{B_i} + \xi_r^{B_i} + \zeta_r^{B_i}\right)\right.$$
$$\left. \times A_1\left(t_k^{B_i} + \xi_k^{B_i} + \zeta_k^{B_i}\right)A_2\left(t_r^{B_i} + \xi_r^{B_i} + \zeta_r^{B_i}\right)X_k^{B_i}Y_r^{B_i}\bigg|B_i\right]$$

$$+2E\left[\sum_{k=1}^{n_{B_i}-1}\sum_{r=k+1}^{n_{B_i}} I_{]i+j-2,i+j-1]}\left(t_k^{B_i} + \xi_k^{B_i} + \zeta_k^{B_i}\right)I_{]i+j-2,i+j-1]}\left(t_r^{B_i} + \xi_r^{B_i} + \zeta_r^{B_i}\right)\right.$$
$$\left. \times A_2\left(t_k^{B_i} + \xi_k^{B_i} + \zeta_k^{B_i}\right)A_1\left(t_r^{B_i} + \xi_r^{B_i} + \zeta_r^{B_i}\right)Y_k^{B_i}X_r^{B_i}\bigg|B_i\right].$$

Finally, the result follows by taking the expectations with respect to conditional distribution of $\zeta_k^{B_i}$ and $\zeta_r^{B_i}$ on $B_i$.

For identity (2), we have

$$E\left[W^2(t)\right] = E\left[\sum_{i=2}^{t}\sum_{j=t+2-i}^{t} W_{i,j}(t)\sum_{k=2}^{t}\sum_{l=t+2-k}^{t} W_{k,l}(t)\right]$$

$$= E\left[\sum_{i=2}^{t}\sum_{j=t+2-i}^{t} W_{i,j}^2(t) + \left\{\sum_{i=2}^{t}\sum_{\substack{j=t+2-i \\ l\neq j}}^{t}\sum_{l=t+2-i}^{t} W_{i,j}(t)W_{i,l}(t) + \sum_{i=2}^{t}\sum_{j=t+2-i}^{t}\sum_{\substack{k=2 \\ k\neq i}}^{t}\sum_{l=t+2-k}^{t} W_{i,j}(t)W_{k,l}(t)\right\}\right]$$

$$= \sum_{i=2}^{t}\sum_{j=t+2-i}^{t} E\left[W_{i,j}^2(t)\right] + \left\{\sum_{i=2}^{t}\sum_{j=t+2-i}^{t}\sum_{\substack{l=t+2-i \\ l\neq j}}^{t} E\left[W_{i,j}(t)W_{i,l}(t)\right]\right.$$

$$\left. + \sum_{i=2}^{t}\sum_{j=t+2-i}^{t}\sum_{\substack{k=2 \\ k\neq i}}^{t}\sum_{l=t+2-k}^{t} E\left[W_{i,j}(t)\right]E\left[W_{k,l}(t)\right]\right\},$$

where, for $l \neq j$,

$$E\left[W_{i,j}(t)W_{i,l}(t)\right] = E\left[\sum_{k=1}^{n_{B_i}} I_{]i+j-2,i+j-1]}\left(t_k^{B_i} + \xi_k^{B_i} + \zeta_k^{B_i}\right)\left\{A_1\left(t_k^{B_i} + \xi_k^{B_i} + \zeta_k^{B_i}\right)X_k^{B_i} + A_2\left(t_k^{B_i} + \xi_k^{B_i} + \zeta_k^{B_i}\right)Y_k^{B_i}\right\}\right.$$
$$\left.\times \sum_{r=1}^{n_{B_i}} I_{]i+l-2,i+l-1]}\left(t_r^{B_i} + \xi_r^{B_i} + \zeta_r^{B_i}\right)\left\{A_1\left(t_r^{B_i} + \xi_r^{B_i} + \zeta_r^{B_i}\right)X_r^{B_i} + A_2\left(t_r^{B_i} + \xi_r^{B_i} + \zeta_r^{B_i}\right)Y_r^{B_i}\right\}\Big| B_i\right]$$

$$= E\left[\sum_{k=1}^{n_{B_i}-1}\sum_{r=k+1}^{n_{B_i}} I_{]i+j-2,i+j-1]}\left(t_k^{B_i} + \xi_k^{B_i} + \zeta_k^{B_i}\right)\left\{A_1\left(t_k^{B_i} + \xi_k^{B_i} + \zeta_k^{B_i}\right)X_k^{B_i} + A_2\left(t_k^{B_i} + \xi_k^{B_i} + \zeta_k^{B_i}\right)Y_k^{B_i}\right\}\right.$$
$$\times I_{]i+l-2,i+l-1]}\left(t_r^{B_i} + \xi_r^{B_i} + \zeta_r^{B_i}\right)\left\{A_1\left(t_r^{B_i} + \xi_r^{B_i} + \zeta_r^{B_i}\right)X_r^{B_i} + A_2\left(t_r^{B_i} + \xi_r^{B_i} + \zeta_r^{B_i}\right)Y_r^{B_i}\right\}\Big| B_i$$
$$+ \sum_{r=1}^{n_{B_i}-1}\sum_{k=r+1}^{n_{B_i}} I_{]i+l-2,i+l-1]}\left(t_r^{B_i} + \xi_r^{B_i} + \zeta_r^{B_i}\right)\left\{A_1\left(t_r^{B_i} + \xi_r^{B_i} + \zeta_r^{B_i}\right)X_r^{B_i} + A_2\left(t_r^{B_i} + \xi_r^{B_i} + \zeta_r^{B_i}\right)Y_r^{B_i}\right\}$$
$$\left.\times I_{]i+j-2,i+j-1]}\left(t_k^{B_i} + \xi_k^{B_i} + \zeta_k^{B_i}\right)\left\{A_1\left(t_k^{B_i} + \xi_k^{B_i} + \zeta_k^{B_i}\right)X_k^{B_i} + A_2\left(t_k^{B_i} + \xi_k^{B_i} + \zeta_k^{B_i}\right)Y_k^{B_i}\right\}\Big| B_i\right]$$

$$= \sum_{k=1}^{n_{B_i}-1} \int_{i+j-2-t_k^{B_i}-\xi_k^{B_i}}^{i+j-1-t_k^{B_i}-\xi_k^{B_i}} \left\{A_1\left(t_k^{B_i} + \xi_k^{B_i} + v\right)E\left[X_k^{B_i}\big|\zeta_k^{B_i}=v\right] + A_2\left(t_k^{B_i} + \xi_k^{B_i} + v\right)E\left[Y_k^{B_i}\big|\zeta_k^{B_i}=v\right]\right\} dF_{\zeta_k^{B_i}|B_i}(v)$$

$$\times \sum_{r=k+1}^{n_{B_i}} \int_{i+l-2-t_r^{B_i}-\xi_r^{B_i}}^{i+l-1-t_r^{B_i}-\xi_r^{B_i}} \left\{A_1\left(t_r^{B_i} + \xi_r^{B_i} + w\right)E\left[X_r^{B_i}\big|\zeta_r^{B_i}=w\right] + A_2\left(t_r^{B_i} + \xi_r^{B_i} + w\right)E\left[Y_r^{B_i}\big|\zeta_r^{B_i}=w\right]\right\} dF_{\zeta_r^{B_i}|B_i}(w)$$

$$+ \sum_{r=1}^{n_{B_i}-1} \int_{i+l-2-t_r^{B_i}-\xi_r^{B_i}}^{i+l-1-t_r^{B_i}-\xi_r^{B_i}} \left\{A_1\left(t_r^{B_i} + \xi_r^{B_i} + v\right)E\left[X_r^{B_i}\big|\zeta_r^{B_i}=v\right] + A_2\left(t_r^{B_i} + \xi_r^{B_i} + v\right)E\left[Y_r^{B_i}\big|\zeta_r^{B_i}=v\right]\right\} dF_{\zeta_r^{B_i}|B_i}(v)$$

$$\times \sum_{k=r+1}^{n_{B_i}} \int_{i+j-2-t_k^{B_i}-\xi_k^{B_i}}^{i+j-1-t_k^{B_i}-\xi_k^{B_i}} \left\{A_1\left(t_k^{B_i} + \xi_k^{B_i} + w\right)E\left[X_k^{B_i}\big|\zeta_k^{B_i}=w\right] + A_2\left(t_k^{B_i} + \xi_k^{B_i} + w\right)E\left[Y_k^{B_i}\big|\zeta_k^{B_i}=w\right]\right\} dF_{\zeta_k^{B_i}|B_i}(w)$$

**Appendix C**

**- Conditional ATRP with parameters uncertainty**

- **Step 1:** We observe the original upper triangle $(X_k, Y_k, T_k, \xi_k, \zeta_k)$;

- **Step 2:** We estimate the parameters of the distributions of $\xi_k$ and $\zeta_k$ using maximum likelihood;

- **Step 3:** We estimate the parameters $\alpha_1$ and $\alpha_2$ using a quasi-likelihood based on the Poisson distribution;

- **Step 4:** Repeat 100 000 times the following :

  a. Using the asymptotic multivariate normal distribution of the log-likelihood of $\xi_k$ and $\zeta_k$, simulate the parameters of the distribution of $\zeta_k$;
  b. Using the asymptotic multivariate normal distribution of the quasi-likelihood estimated in Step 3, simulate values of $\alpha_1$ and $\alpha_2$;
  c. Using the simulated values of $\alpha_1$ and $\alpha_2$, calculate deflated values of $X_k$ and $Y_k$ which we define as $\tilde{X}_k$ and $\tilde{Y}_k$ respectively;
  d. Resample the $\tilde{X}_k$ and $\tilde{Y}_k$ to obtain $\tilde{X}_k^*$ and $\tilde{Y}_k^*$;
  e. Using the expected maximization algorithm, estimate the parameters of the distribution of the indemnities and expenses using $\tilde{X}_k^*$ and $\tilde{Y}_k^*$.